\documentclass[12pt]{article}
\usepackage{epsfig,a4}
\usepackage{cite}
 \advance\oddsidemargin by -1in
 \advance\evensidemargin by -1in
 \advance\topmargin by -0.0in
 \makeindex
\usepackage{graphics}

 \newcommand\la{\langle}
 \newcommand\ra{\rangle}
 \newcommand\beq{\begin{equation}}
 
 \newcommand\eeq{\end{equation}}
 \newcommand\beqn{\begin{eqnarray}}
 \newcommand\eeqn{\end{eqnarray}}

 \def\BA{\begin{eqnarray}}
 \def\BE{\begin{equation}}
 \def\BF{\begin{figure}[htb]}
 \def\BT{\begin{table}[htb]}
 \def\EA{\end{eqnarray}}
 \def\EE{\end{equation}}
 \def\EF{\end{figure}}
 \def\ET{\end{table}}

 \def\la{\langle}
 \def\ra{\rangle}
 \def\mb{\,\mbox{mb}}
 \def\fm{\,\mbox{fm}}
 \def\GeV{\,\mbox{GeV}}
 \def\MeV{\,\mbox{MeV}}

 \def\lsim{\mathrel{\rlap{\lower4pt\hbox{\hskip1pt$\sim$}}
    \raise1pt\hbox{$<$}}}
\def\gsim{\mathrel{\rlap{\lower4pt\hbox{\hskip1pt$\sim$}}
    \raise1pt\hbox{$>$}}} 
\begin{document}
\date{}

 \title{\bf Nuclear Hadronization: Within or 
Without?\footnote{Based on
talks given by B.Z.K. at the Fourth International Conference on
Perspectives in Hadronic Physics, Trieste, Italy, May 12-16, 2003; and at
the EuroConference on Hadron Structure Viewed with Electromagnetic Probes,
Santorini, Greece, October 7-12, 2003.} } 

\maketitle

\begin{center}

\vspace*{-1.5cm}

{\large B.Z.~Kopeliovich$^{1-3}$, 
J.~Nemchik$^4$,
E.~Predazzi$^5$ and A.~Hayashigaki$^2$
}\\[0.5cm]
{$^{1}$Max-Planck-Institut f\"ur Kernphysik, 
Postfach 103980, 69029 Heidelberg, Germany \\
$^{2}$Institut f\"ur Theoretische Physik der Universit\"at, 93040
Regensburg, Germany\\
$^{3}$Joint Institute for Nuclear Research, Dubna, 141980 Moscow
Region, Russia\\
$^{4}$Institute of Experimental Physics SAS, Watsonova 47,
CS-04353 Kosice, Slovakia\\
$^{5}$Dipartimento di Fisica Teorica, Universit\`a di Torino
and INFN, Sezione di Torino,\\ I-10125, Torino, Italy
}

\end{center}

\vspace{1cm}

\abstract{Nuclei are unique analyzers for the early stage of the
space-time development of hadronization. DIS at medium energies is
especially suitable for this task being sensitive to hadronization
dynamics, since the production length is comparable with the nuclear size.
This was the driving motivation to propose measurements at HERMES using
nuclear targets, and to provide predictions based on a pQCD model of
hadronization \cite{knp}. Now when the first results of the experiment are
released \cite{hermes,hermes-q2}, one can compare the predictions with the
data. The model successfully describes with no adjustment the nuclear
effects for various energies, $z_h$, $p_T$, and $Q^2$, for different
flavors and different nuclei. It turns out that the main source of nuclear
suppression of the hadron production rate is attenuation of colorless
pre-hadrons in the medium. An alternative model \cite{ww} is based upon an
ad hoc assumption that the colorless pre-hadron is produced outside the
nucleus. This model has apparent problems attempting to explain certain
features of the results from HERMES.  A good understanding of the
hadronization dynamics is important for proper interpretation of the
strong suppression of high-$p_T$ hadrons observed in heavy ion collisions
at RHIC. We demonstrate that the production length is even shorter in this
case and keeps contracting with rising $p_T$.\\[0.5cm]
 }

\newpage

\section{Introduction}\label{intro} 

Recent measurements by the HERMES collaboration \cite{hermes} of
semi-inclusive production of hadrons in deep-inelastic scattering (DIS)
off nuclei have provided precious information about the space-time
development of hadronization. These measurements were proposed \cite{knp}
back in 1995 and predictions were made within a model of hadronization
based on perturbative QCD. One of the goals was to reach a better
understanding of in-medium hadronization in order to provide a more
reliable interpretation of high-$p_T$ hadron production in heavy ion
collisions considered as a probe for the dense matter produced.

Now, when some of the results of these measurements are released, it is
proper to compare the predictions with the data and draw conclusions.  In
the present paper we revisit the problem of in-medium hadronization,
re-introduce the model, and perform calculations for HERMES kinematics. We
demonstrate that absorption of the produced colorless pre-hadron is the
main source of nuclear suppression. Our predictions need no adjustment and
are in good agreement with the data.

Older models \cite{bialas,kn,bc,k90,krakow} were based on the string
model which were quite pedagogical and helpful for intuitive understanding 
of the space-time development of hadronization. On the other hand, one
could not consider such a nonperturbative phenomenology as a realistic
scheme for hard reactions like DIS. These models had a very little
predictive power, no access to $Q^2$ and $p_T$-dependences, missed
the effect of color transparency, etc.

The release of data from the HERMES experiment has stimulated a new wave
of theoretical models. All of them have low predictive power and are
fitted to data which are supposed to be explained. This is why all
of them agree with data, although they explore quite different physical
ideas.

Some of the models \cite{amp,giessen,akopov} continue
developing already known ideas based on the phenomenology of
nonperturbative hadronization. Others \cite{ww,arleo}, employ the idea of
perturbative induced energy loss, but push to the
extreme. Namely, they make two strong ad hoc assumptions which have no
justification. First, it is assumed that pre-hadrons are always produced 
outside the nucleus. Second, the effect of induced energy loss
is accounted for via a simple shift of the variable in the fragmentation
function, which is equivalent to the assumption that hadronization starts
only after the leading parton leaves the nucleus.

Since the induced energy loss scenario is a popular model or the
nuclear suppression of high-$p_T$ events observed at RHIC, we confront
some of the predictions of this model with data from HERMES. We sort out
those observables which are sensitive to the model assumptions.

\subsection{Why nuclear target?}\label{subintro1}

 Particles produced in hadronic collisions and detected at macroscopic
distances carry limited information about the hadronization dynamics. The
most important details are hidden at the early stage. Nuclear targets
provide a unique opportunity to look at the early stage of hadronization
at distances of a few Fermi from the origin. A quark-gluon system
originated from DIS propagates through the nuclear medium and interacts
with other nucleons. Modification of the differential cross section of
particle production can bring forth precious information about the
structure of the excited system and its space-time development.

We assume in what follows that Bjorken $x$ is sufficiently large, $x\sim
0.1$, therefore the lepton interacts incoherently with only one bound
nucleon, no shadowing is possible.  Besides, this region is dominated by
valence quarks, i.e.  one can treat the DIS as electron-quark scattering
with energy transfer, $\nu$, to the knocked out quark.  At small $x$
dominated by the sea the process of DIS looks different (in the nuclear
rest frame): the virtual photon produces two jets, $q$ and $\bar q$, which
share the full energy $\nu$. Therefore, the variable $z_h$, which is the
fraction of $\nu$ transferred to the detected hadron, cannot reach the
kinematic limit $z_h=1$ for either of the two jets, unless one of them
takes the whole energy of the photon. This fact makes the ratio of nuclear 
to nucleon cross sections [see Eq.~(\ref{40})]
fall faster at $z_h\to1$ than at large $x$. Additionally, negative kaons
which at large $x$ are produced mainly from gluons, at small $x$ are
generated by the same mechanism as positive kaons. Thus, one should be
cautious about the range of $x$ involved in the analysis, and make a
proper $x$-binning of DIS data.

One can also use the process of hadronization as a tool for probing the
medium properties. This is the driving idea of the so called "jet
quenching" probe for creation of dense matter in relativistic heavy ion
collisions \cite{miklos}. This tool, however, can work properly only if
the hadronization dynamics is reliably understood, which seems to be still
a challenge.

\subsection{Vacuum and induced energy losses}\label{subintro2}

Energy loss is a hot topic nowadays, however, related discussions are
confusing sometimes, since vacuum and induced energy losses are mixed up.
Due to confinement, an energetic parton cannot propagate in vacuum as a
free particle, but hadronizes and produces a jet of hadrons. The parton
shares its energy with the produced hadrons, therefore it is gradually
losing energy. We call this {\it vacuum energy loss}. As a guidance from
string model \cite{cnn} the rate of energy loss should be constant, of the
order of the string tension $\kappa=1\GeV/\fm$ (see Sec.~2). Therefore,
the energy loss rises linearly with time or length of the path,
 \beq 
\Delta E_{vac}(L) = \kappa L\ .  
\label{10}
 \eeq

If the parton originates from a hard reaction (DIS, high-$p_T$ process,
etc), gluon bremsstrahlung should be an additional or even the main source
of energy loss. Indeed, as a result of a strong kick the parton should
shake off a part of its color field with transverse frequencies controlled
by the strength of the kick. Apparently, energy loss may be very extensive
and may substantially exceed the static value given by the string model.
In this case an approach based on perturbative QCD should be more
appropriate.

One can trace similarity with energy loss in QED. If a plate of a big
capacitor, which transverse dimensions are much larger than the distance
between the plates, is accelerated, the main contribution to the retarding
force comes from the static field. Indeed, the radiation from the edges of
the capacitor is small and can be neglected. This an analog to a color
string (tube) model which might be a reasonable approximation for soft
processes. Indeed, this case is associated with constituent quarks which
size is of the order of the tube length.

On the contrary, if one accelerates point-like charges, the main
contribution to energy loss should be bremsstrahlung, rather than Coulomb
forces. This would be an analogy for quark jets originating from DIS.

Amazingly, the radiative energy loss turns out to rise linearly with the
path length \cite{feri} like in the string model,
 \beq
\Delta E_{vac}(L) = \frac{2}{3\pi}\,\alpha_s(Q^2)\,Q^2\,L\ .
\label{20}
 \eeq
 This happens because the gluons lose coherence with the source, i.e. are
radiated at different time intervals.

In the case of hadronization inside a medium, an additional source of
energy loss is interaction of partons with the medium. Due to multiple
collisions the parton increases its transverse momentum squared linearly
with the path length, since it performs Brownian motion in the transverse
momentum plane. Experimentally, the broadening of transverse momentum is a
rather small effect. For instance, the broadening measured in the
Drell-Yan reaction on a nucleus as heavy as tungsten is only $\Delta
p_T^2\approx 0.1\GeV^2$. Apparently, the induced energy loss \cite{bdmps}
generated by such a small momentum transfer is a small addition to the
vacuum energy loss originated from a hard process (at least in cold
nuclear matter),
 \beq
\Delta E_{ind}(L)= \frac{3}{8}\alpha_s\,\Delta p_T^2\,L
= {3\over4}\,\alpha_s\,C(E)\,\rho_A\,L^2\ ,
\label{30}
 \eeq
 where $C(E)=d\sigma_{\bar qq}(r_T,E)/dr_T^2\Bigr|_{r_T=0}$
\cite{dhk,jkt} is the derivative of the universal phenomenological dipole
cross section which depends on the transverse dipole separation $r_T$; 
$\rho_A$ is the nuclear density. Of course, application of perturbative 
QCD in this case is questionable.

The intuitive interpretation of these results is rather straightforward.
It takes a time, called the coherence time, to radiate a photon or gluon
which becomes incoherent with the source \cite{knp},
 \beq
t_c=\frac{2E\alpha(1-\alpha)}{k_T^2}\ ,
\label{35}
 \eeq
 where $k_T$ and $\alpha$ are the transverse momentum and fraction of
energy $E$ taken away by the radiated quantum. According to the
Landau-Pomeranchuk principle, radiation at long coherence times does not
resolve the details of interaction, whether it was a single kick, or a few
of them. What matters is the final transverse momentum of the parton.
Since the additional transverse momentum gained by the parton due to
multiple interactions in the nuclear medium is very small, the
corresponding correction (induced) to the energy loss is small as well.

This is why it is not easy to see in data the effects of induced energy
loss. In particular, the analysis of nuclear effects in Drell-Yan reaction
performed in \cite{eloss} provided information about vacuum energy loss.
Indeed, according to the kinematics of the E772/E866 experiments (and
other experiments at lower energies) most of events correspond to a short
coherence time for the Drell-Yan process which occurs nearly momentarily
deep inside the nucleus. Prior to that the incoming hadron interacts
softly on the front surface of the nucleus. This collision breaks down the
coherence of the projectile partons which start losing energy on the path
from the soft collision point until the point where the Drell-Yan reaction
takes place. One should expect the vacuum value for the rate of energy
loss which is at least $-dE/dz\approx 1\GeV/\fm$ or higher, much larger
than what one could expect for induced gluon bremsstrahlung. Indeed, the
analyses \cite{eloss} led to $-dE/dz = (2.8 \pm 0.4 \pm 0.5)\GeV/\fm$,
which is a summed effect of vacuum and induced energy losses.

Naively, one might expect that the effects of vacuum energy loss cancel
out in the nucleus to nucleon ratio, since they are identical in the
numerator and denominator. This is not correct, firstly, induced energy
loss is a part of hadronization and it does not result in a simple shift
of the variable in the fragmentation function. Secondly, the vacuum energy
loss controls the time scale of hadron production which has a great impact
on nuclear effects (see next section).

\subsection{Production and formation times}\label{subintro3}

 One should discriminate between formation of the final hadron and
production of a colorless state which is not yet an eigenstate of the mass
matrix and may be projected to various hadronic wave functions. We call
such states pre-hadrons and the related time scale {\it production time},
$t_p$.  The properties of such pre-hadron states and their attenuation in
nuclei have been intensively studied over the last two decades, both
theoretically and experimentally (see e.g. in \cite{pr}). It has been
proven that such pre-hadrons of a reduced size are indeed produced, and
that nuclei are indeed more transparent for such states. A signal of color
transparency in diffractive DIS was observed recently in the HERMES
experiment \cite{borissov,knst}.

To form the hadronic wave function the constituents of the pre-hadron have
to circle at least once along their orbits. Dilated by Lorentz
transformation this time scale, which we call {\it formation time}, $t_f$,
is proportional to $z_h$.

 On the other hand, the production time $t_p$ was
proven in \cite{kl,kn} and confirmed within the LUND model in \cite{bg} to
vanish proportionally to $1-z_h$ at $z_h\to 1$. The latter property
follows from energy conservation: the longer the hadronization process is
lasting, the more energy is lost by the leading quark, unless its energy
falls below the desired hadron energy. In particular, at $z_h=1$ no energy
loss for hadronization is allowed, so the colorless ejectile has to be
created instantaneously.

Note that an averaged hadronization process is lasting a long time
proportional to the energy of the parton initiated the jet. However, the
mean energy fractions $z_h$ of the produced partons are quite small.
Selecting rare events with a leading hadron produced with large $z_h$ one
imposes a severe restriction upon the hadron production time which must be
proportional to $(1-z_h)$.

\subsection{What are the observables?}\label{subintro4}

The nuclear medium affects the momentum distribution of the produced
particles, which is called sometimes a modified fragmentation function
$D^A_{eff}(z_h,p_T,Q^2,\nu)$, where $p_T$ is the transverse momentum of
the hadron, and $\nu$ and $Q^2$ are the energy and virtuality of the
photon respectively. The energy dependence of $D^A_{eff}$ signals that QCD
factorization is broken and $D^A_{eff}$ cannot be treated as a
fragmentation function, e.g. the QCD evolution equations cannot be
applied. Only at very large $\nu$ and $Q^2$ nuclear effects disappear and
factorization is restored. In this trivial limit, however, all physical
information we are interested in is missed.

Experimental results are usually presented in a simple form, as a ratio
of the nucleus-to-nucleon hadron multiplicities,
 \beq
R_A(z_h,p_T,Q^2,\nu)=\frac{dn(\gamma^*A\to hX)/dz_h d^2p_T}
{A\, dn(\gamma^*N\to hX)/dz_h d^2p_T}\ .
\label{40}
 \eeq
 Because of limited statistics this four-dimensional ratio is usually
presented as one-dimensional ratios integrated over other variables. Some
of correlations are, however, very informative. For instance, the $p_T$
dependence of the ratio binned in $z_h$ brings forth precious information
about the space-time pattern of hadronization (see Sec.~9.3).

Nuclear modification of the hadronic spectra was considered for high-$p_T$
hadron production in \cite{kn,kk}, for DIS in \cite{bialas,bc,krakow} and
for hadroproduction of leading particles on nuclei in \cite{bg,kln}.

\subsection{What can we learn?}\label{subintro5}

 The questions, which can be answered in such an analysis 
are:\\[0.1cm]
 ~--~ {\it How long does it take to produce a pre-hadron?}\\ Nuclear
suppression of hadron production, especially its $z_h$ and $p_T$
dependences, are sensitive to the production time $t_p$. \\[0.1cm]
 ~--~{\it How does the produced pre-hadron evolve and
attenuate in nuclear matter?}	\\
 Since we are interested in production of leading particles, we should
forbid inelastic interaction of the produced colorless wave packet at
times $t > t_p$. It would lead to a new hadronization process which ends
up with a smaller value of $z_h$. The corresponding correction is easy to
estimate and it turns out to be small. The condition of no interaction
results in an attenuation of the produced wave packet in nuclear matter,
but the absorption cross section of the pre-hadron may be different from
the hadronic one. Particularly, in DIS with high photon virtuality $Q^2$
the produced wave packet may have a smaller transverse size, similar to
what is known for exclusive particle production. Correspondingly, the
color transparency effects \cite{zkl,bbgg} should be important and we
take care of that in what follows.\\[0.1cm]
 ~--~{\it Does a fast quark attenuate in nuclear matter?}\\
 At first glance an energetic quark should not attenuate since it cannot
be stopped or absorbed in the medium. Multiple soft interactions can only
rotate the quark in color space, while the energy loss is a small fraction
of the quark energy. However, by ``attenuation'' of a quark, we mean the
suppression of the production rate of the final hadron produced with a
certain momentum, related to the quark interactions in the nuclear medium.  
In the string model any color-exchange interaction leads to formation of a
new string, but with reduced initial energy. This fact results in a
reduction of hadron production at large $z_h\to 1$ \cite{k90}. Moreover,
the leading quark may pick up new strings due to presence of higher Fock
components, giving rise to induced energy loss in this model \cite{eloss}.
If one tries to treat the reduced production rate of hadrons as an
effective attenuation of the quark in the medium, the effective quark
absorption cross section cannot be treated as an universal parameter. Its
value essentially varies with geometry and kinematics \cite{k90}. In the
present paper we include induced energy loss explicitly.

\section{Lessons from the string model}\label{srting}

Apparently, the nonperturbative string model cannot be a realistic
approach to such a hard process as DIS. The model does not have any
access to $Q^2$-dependence, color transparency, etc., which are just the 
phenomena that are 
crucial for understanding the properties of in-medium hadronization. At
the same time, the model is simple, very intuitive, and helps to
understand the general features of the space-time development of
hadronization.

As was already mentioned, we assumed that data are taken at large Bjorken
$x>0.1$ to make sure that valence quarks in the target nucleon give the
main contribution.  A color string\footnote{Of course a one-dimensional
string is an idealization.  We neglect the transverse size of the
color-flux tube \cite{cnn}, unless otherwise specified.} formed between
the quark and the debris of the target nucleon, slowing down the former
and speeding up the latter.  Naively, one may think about a long string
stretched across the nucleus.  Instead, a very short object is propagating
through the nucleus.  Indeed, a simple kinematics shows that the maximal
possible length of the string in the rest frame of the nucleus is,
 \beq
L_{max}=\frac{m}{\kappa}\ ,
\label{2}
 \eeq
 where $m$ is the mass of the debris of the target nucleon (e.g. a
diquark) attached to the slow end of the string.  The string tension
$\kappa=1/(2\pi\alpha_R^\prime)\approx 1\,$GeV/fm is related to the slope
$\alpha^\prime_R$ of the Regge trajectories \cite{cnn}.  Since the
nucleon debris are probably lighter than the nucleon, $L_{max} <
1\,$fm.  In
reality the string should be even much shorter due to the Schwinger
phenomenon, spontaneous $q\bar{q}$ pair production from vacuum breaking
the string to shorter pieces.

Thus, a heavy but short string with the effective mass $M\approx
\sqrt{s}=\sqrt{m_N^2-Q^2+2m_N\nu}$ propagates through the nucleus. Due to
$q\bar{q}$ pair production in the color field of the string light pieces
of the string are chipped off, while the leading heavy part of the
string keeps propagating on. Apparently, the leading quark is slowing
down, losing energy with a constant rate,
 \beq
\frac{d\,E}{d\,t}= -\,\kappa\ ,
\label{3}
 \eeq
 where $E(t)$ is the quark energy and $t=z$ are the time and longitudinal
coordinate. Note that the rate of energy loss (\ref{3}) is a constant
which is independent of the quark flavor, energy and virtuality, and is
invariant relative to longitudinal Lorentz boosts.  Correspondingly, the
total energy loss rises linearly with the length $\Delta z$ of the path,
$\Delta E=\kappa \Delta z$.  The lost energy goes to acceleration of the
target debris and production of new hadrons. The process of hadronization
completes when the mass of the leading piece of the string is reduced by
many decays down to the hadronic mass scale and the leading pre-hadron
(or a cluster) is produced.

How does the production time depend on the initial quark energy $\nu$ and
its fraction $z_h$ carried by the produced hadron? For low energy hadrons
it is short, but rises with their energy, $t_{p} \approx \nu
z_h/2\kappa$. One might think that such a linear $z_h$-dependence is an
unquestionable truth since it is dictated by Lorentz time dilation. It
was claimed, however, in \cite{kl,kn} and confirmed within Lund model
\cite{bg}, that there is an opposite trend for leading hadrons, namely,
the production time vanishes like $t_p =(1-z_h)\nu/\kappa$ at $z_h\to1$.
Although it was first derived in the string model \cite{kn,kl}, this is a
general property, since is dictated by conservation of energy. Indeed,
unless color neutralization happens, i.e. a colorless system with energy
$z_h\nu$ is produced, the leading quark keeps losing energy. Therefore,
the smaller $(1-z_h)$ is, the faster this process must be completed,
otherwise the energy needed for the leading hadron production will be
wasted.

Note, that the produced colorless object should be treated as a
pre-meson. It may take a long time proportional to $z_h\nu$ to form the
final hadron wave function. We call this formation time, and it is always
longer than the production time, $t_f > t_p$.

Thus, we arrive at two different types of end-point behavior
of the production time \cite{kl,kn,bg},
 \beq
t_{p} = \left\{
\begin{array}{c}
\frac{\nu}{2\kappa}\,z_h\hspace{2.2cm} z_h\ll 1\\
{\nu\over\kappa}\,(1-z_h)\hspace{1cm} 1-z_h\ll 1
\end{array}\right.
\label{4}
 \eeq

 The concept of production time is crucial in the case of a nuclear
target:  a colorless pre-hadron produced inside the nucleus should not
interact (inelastically) on the way out of the nucleus, otherwise
hadronization will be triggered, and the final hadron will emerge with a
substantially reduced energy.  This restriction leads to an attenuation
which depends on the interaction cross section of the pre-hadron (see
below).  For this reason different processes on nuclei are suppressed near
the kinematic limit $z_h \to 1$, for example, back-to-back high-$p_T$
dihadron production, DIS, forward particle production in hadron-nucleus
interactions, etc.

As was mentioned above, the production time Eq.~(\ref{4}) corresponds to
creation of a fast colorless piece of the string, a pre-hadron, rather
than the final hadron.  The latter is supposed to have a wave function
describing the specific distribution of quark momenta inside the hadron,
different from the quark momentum ordering in the string.  The wave
function of the hadron is developing over a longer period of time
(formation time), which is proportional to the hadron energy, $z_h\nu$,
rather than to $(1-z_h)\nu$. These two time scales are mixed up sometimes.
At low $Q^2$ the pre-hadron has a large absorption cross section, and
nuclear attenuation is controlled mainly by the production time
Eq.~(\ref{4}). In this case the formation time has a moderate influence on
nuclear effects. On the contrary, at large $Q^2$ the formation time scale
is more important for nuclear attenuation. We will include both effects in
our calculations.

Another question which may be answered within the string model is how
multiple interactions of the hadronizing quark at earlier stage, $t <
t_p$ affects the hadron production rate? Sometimes \cite{bc,hermes}
this attenuation is treated as absorption with an effective quark-nucleon
cross section treated as a universal fitted parameter. However, the
physical origin of such an attenuation discussed in \cite{k90} shows that
it cannot be described in terms of a universal parameter like a quark
absorption cross section. Indeed, every interaction of the leading quark
leads to a new hadronization process starting from the very beginning,
but with a reduced initial energy $E_q(z)=\nu-\kappa z$, where $z$ is the
distance covered by the quark. Since the energy of the final hadron is
fixed, the variable of the fragmentation function should be redefined as,
$z_h \Rightarrow z_h/(1-\kappa z/\nu)$. Apparently, the increase of $z_h$
should lead to a suppression which depends on $\nu$ and $z_h$ and 
cannot
be treated as a universal parameter \cite{k90}.

Moreover, multiple quark interactions at $t<t_p$ induce an additional
energy loss with a rate rising linearly with the path length $z$
\cite{eloss},
 \beq
\frac{d\,E}{d\,z}=
- \kappa\Bigl[1 + \la n(z)\ra\Bigr]\ ,
\label{5}
 \eeq
 where $\la n(z)\ra = \sigma^{qN}_{in}\,\rho_A\,z$ is the mean number of
collisions experienced by the quark over the distance $z$;
$\sigma^{qN}_{in}$ is the model dependent effective quark-nucleon
inelastic cross section. The first term in (\ref{5}), same as in
(\ref{3}), is the usual retarding force produced by the string, while the
second term is due to additional strings created by multiple interaction
of the quark. These extra terms correspond to higher Fock component of
the quark and can be motivated within the dual parton model \cite{dpm}
(more details can be found in \cite{eloss}).

Thus, the total energy loss on a distance $L$ contains a correction
$\propto L^2$.
 \beq
\Delta E(L) = \kappa~L\bigl [1 +
\frac{1}{2}\sigma^{qN}_{in}\rho_{A}~L\bigr ]\, .
\label{37}
 \eeq
 In the additive quark model $\sigma^{qN}_{in}\sim
10$\,mb, which is the same value as follows from the dual parton model
\cite{dpm}. Note, that this value is twice as large as in the more
realistic model \cite{k3p}, which demonstrates that the dominant part
of the hadronic cross section does not obey quark additivity.

It is questionable whether one can apply this phenomenology to DIS. At
this point we feel that our unjustified assumption that soft interaction
phenomenology, like the string model, is relevant for final state
interaction in DIS has been pushed to the extreme, and we should consider
a more realistic approach based on perturbative QCD.

\section{Radiative energy loss: hadronization in\\ vacuum}

\subsection{Vacuum energy loss}\label{3.1}

When a high-momentum parton is produced in a hard reaction, it triggers
gluon bremsstrahlung, which lasts until the moment of color
neutralization. Gluons are not radiated instantaneously, but it takes
time to lose coherence with the parent parton. This quantum-mechanical
uncertainty called coherence time can be introduced differently.  A
strict definition comes within precise calculations and is related to
interference between quanta radiated at different positions. For
instance, two amplitudes of gluon radiation by a quark at different
coordinates separated by the longitudinal distance $\Delta z$ get phase
shift $\Delta z q_L$, where $q_L = (M_{Gq}^2-m_q^2)/2E_q$ is a difference
between the longitudinal momenta of the quark and the produced
quark-gluon final state which has effective mass squared,
 \beq
M_{Gq}^2 = \frac{k_T^2+\alpha m_q^2}
{\alpha(1-\alpha)}\,.
\label{50}
 \eeq
 Here $\vec k_T$ and $\alpha$ are the transverse momentum and fraction of
the light-cone momentum of the quark carried by the gluon; $E_q=\nu$ is
the initial quark energy (we assume that Bjorken $x$ is large and the
quark gets the whole energy transfer).

Apparently, only those gluons can interfere, which are radiated within the
coherence time interval, or coherence length (we assume $z=t$ and neglect
the quark mass hereafter), given by Eq.~(\ref{35}) with $E=\nu$. This time
may be also interpreted as the lifetime of the quark-gluon fluctuation, or
as the formation time of radiation. At any rate, only those gluons may be
treated as radiated during time $t$, which have coherence time $t_c < t$,
otherwise they are still coherent with the quark and should be considered
as a part of its color field.

Correspondingly, one can estimate the energy deposited into the radiation
as
 \beq
\Delta E(t)=\nu\
\int\limits_{\lambda^2}^{Q^2} dk_T^2
\int\limits_0^1 d\alpha\,\alpha\,\frac{dn_{G}}{dk^2_Td\alpha}\,
\Theta(t-t_c)
\label{70}
 \eeq
 where $\lambda$ is an infrared cutoff which we fix at
$\lambda=\Lambda_{QCD}=200\MeV$. The number of radiated gluons
$dn_{G}/d\alpha dk_T^2 = \gamma/\alpha k_T^2$, where $\gamma
=4\alpha_s(k_T^2)/3\pi$ \cite{gb,feri}. As usual, we employ the
approximation of independent and soft radiation, $\alpha\ll 1$.

Eq.~(\ref{70}) leads to a linear time dependence of energy loss,
 \beq
\Delta E(t) = {\gamma\over2}\,(Q^2-\lambda^2)\,t 
\label{80}
 \eeq
 i.e. the rate of energy loss is constant, as it is for strings. Although
gluons are continuously radiated with different coherent times and
different energies, the amount of energy taken away per unit of time
remains the same. This interesting observation was demonstrated in
\cite{feri,bh}.

The radiated energy is distributed among the produced hadrons, but we are
interested in the most energetic one, the leading hadron, which carries
the largest fraction $z_h$ of the initial momentum of the quark. This
condition and energy conservation impose additional restrictions on the
integration in Eq.~(\ref{70}), resulting in a different time development
of the hadronization and the energy loss.

If the leading hadron is produced with large $z_h$ and includes the
leading quark, none of the radiated gluon may leave with a fraction of
the initial quark momentum exceeding $\alpha > 1-z_h$. For instance, in
the limiting case $z_h \to 1$ any radiation is forbidden, which leads to a
so-called Sudakov's suppression of the cross section. With decreasing
$z_h$, this restriction gradually softens, but still should not be
forgotten.

In order to estimate the time-dependence of radiative energy loss
specified for
the case of the leading hadron production we introduce energy
conservation in Eq.~(\ref{70}) via a step-function $\Theta(1-z_h-\alpha)$.
The result of integration reads,
 \beqn
&&\Delta E(t) = \frac{\gamma}{2}\,t\,
(Q^2 - \lambda^2)\,\Theta(t_1-t)
\nonumber\\  & + & 
\left\{ \nu\gamma (1 - z_h)\left[1 + 
\ln\left(\frac{t}{t_1}\right)\right] 
-\frac{\gamma}{2}\lambda^2 t \right\}
\Theta(t-t_1)\Theta(t_2-t)
\nonumber\\  & + & \nu\gamma (1 -
z_h)\ln\left(\frac{Q^2}{\lambda^2}\right)\Theta(t-t_2)\,
\label{90}
 \eeqn
 where 
 \beq
  t_1 = \frac{2\nu}{Q^2}(1-z_h) = \frac{1-z_h}{x\,m_N}\ ,
   \label{100}
 \eeq
 \beq
  t_2={Q^2\over \lambda^2}t_1\,,
\label{110}
 \eeq
 and $x=Q^2/2m_N\nu$ is Bjorken variable.

\begin{figure}[tbh]
\centerline{\rotatebox{270}{\psfig{figure=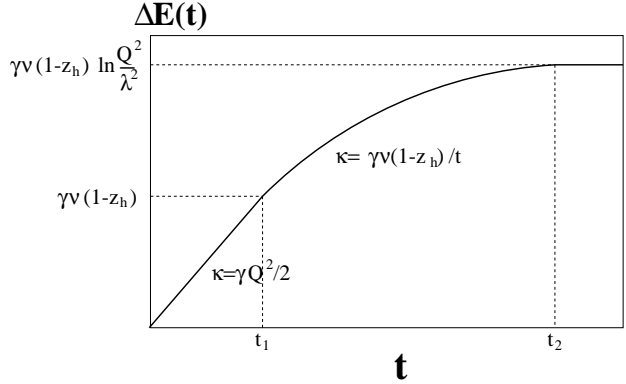,width=5cm}}}
\begin{center}\parbox{13cm}{
 \protect\caption{Time-dependence of energy loss corresponding to
Eq.~(\ref{90}). The notations are explained in the text.}
 \label{eloss} 
}\end{center} 
 \end{figure}

The time dependence of radiative energy loss is illustrated in
Fig.~\ref{eloss} assuming $Q^2\gg\lambda^2$. During the time interval $t
< t_1$, the rate of energy loss is constant, $-dE/dt=\gamma Q^2/2$,
exactly as in the case of Eqs.~(\ref{70}), (\ref{80}) with no restriction
on the radiated energy.

Over a longer time interval, more energetic gluons can be emitted in
accordance with Eq.~(\ref{35}) and the restriction $\alpha < 1-z_h$
becomes effective. As a result, the loss of energy slows down to a
logarithmic $t$-dependence. This implies that the rate of energy loss
decreases as function of time as $dE/dt=-\gamma \nu(1-z_h)/t$.

\subsection{Sudakov's suppression}

According to Eq.~(\ref{90}) at much longer times either very energetic
gluons can be emitted (which may be forbidden by the condition $\alpha <
1-z_h$), or gluons with very small $k_T$. The latter are cut off in
Eq.~(\ref{70}). This completely eliminates any energy loss at long times
$t > t_2$ as one can see from Eq.~(\ref{90}) (see also Fig.~\ref{eloss}).
This sounds puzzling, unless the suppression due to a Sudakov-type factor
is taken into account. This suppression becomes active $t > t_1$, when the
production of the leading hadron with large $z_h$ restricts the spectrum
of radiation.

Summing up all the nonradiated gluons we get the following Sudakov's
suppression factor,
 \beq
S(t,z_{h},Q^2,\nu) = \exp\Bigl[-\tilde{n}_{G}(t,z_h,Q^2,\nu)\Bigr]\ ,
\label{150}
 \eeq  
 which corresponds to the Poisson distribution of independently radiated
soft gluons.  The number of gluons which have sufficiently short coherence
time to be radiated during time interval $t$, but conflicting with the
leading hadron production since they are too energetic, $\alpha > 1-z_h$,
is given by,
 \beq
\tilde{n}_{G}(t,z_h,Q^2,\nu) =
\gamma \int\limits_{1-z_h}^1
\frac{d\alpha}{\alpha}
\int\limits_{\lambda^2}^{Q^2}\frac{dk_T^2}{k_T^2}
[1 - \exp(-t/t_c)]\ .
\label{155}
 \eeq
 To be more realistic we smoothed out the step-function $\Theta(t-t_c)$
here, as well as in (\ref{90}). We assumed that the radiation rate of
gluons with coherence time $t_c$ has a decay rate distribution,
 \beq
\frac{dP(t)}{dt} = \frac{1}{t_{c}}\,e^{-t/t_{c}}\ .
   \label{120}
 \eeq
Then the probability to radiate a gluon during time interval
$t$ is,
 \beq
P(t) = 1 - e^{-t/t_{c}}\, ,
\label{130}
 \eeq
 which is a replacement for the step-function $\Theta(t-t_c)$ in
(\ref{90}) and (\ref{155}).

The Sudakov formfactor, Eq.~(\ref{150}), substantially reduces the mean
time
interval of hadronization and the production time of the leading
pre-hadron. 

\subsection{Leading hadron production}\label{3.2}

The estimate Eq.~(\ref{10}) for energy loss was obtained within the
string model, which has no access to $Q^2$ dependence and assumes that
the final hadron is immediately produced once the quark energy degrades
down to $z_h\nu$. Perturbative QCD gives the opportunity to develop a
more detailed and profound model for hadronization and color
neutralization.  In the large $N_c$ limit each radiated gluon is
equivalent to a $q\bar{q}$ pair, and the gluon bremsstrahlung can be
seen as production of a system of colorless dipoles (compare with
\cite{m,lund}). This would be a nonperturbative mechanism of
hadronization. It is more appropriate to treat hadronization, i.e. the
transition $G\to\bar qq$, perturbatively if the production time is
short. Then it is natural to assume that the leading hadron originates
from the color dipole which includes the leading quark and the
antiquark from the last radiated gluon. In the large $N_c$
approximation they are automatically tuned in color, i.e. this pair is
colorless. This $q\bar{q}$ dipole is to be projected on the hadron wave
function, $\Psi_h(\beta,l_{T})$, where $\beta$ is the fractions of the
hadron light-cone momentum carried by one of the quarks, and $l_T$ is
the transverse momentum of the quarks in the hadron.

Let us evaluate the distribution function
$W(t,z_h,Q^2,\nu)$ for production time $t$ of a pre-meson,
 \beqn
&& W(t,z_h,Q^2,\nu) = N\
\int \limits_0^1 \frac{d\alpha}{\alpha}\ \delta \left
[z_h-\left (1-{\alpha\over 2}\right)
\frac{E_q(t)}{\nu}\right ]
\nonumber\\
&\times& 
\int\limits_{\Lambda^2}^{Q^2} \frac{dk_T^2}{k_T^2}\
\frac{\exp(-t/t_c)}{t_c}\
\int dl_t^2\ \delta \left [l_T^2-{9\over 16}k_T^2 \right ]
\nonumber\\
&\times& 
\int\limits_0^1 d\beta
\ \delta\left [\beta-\frac{\alpha}{2-\alpha} \right ]
|\Psi_h(\beta,l_T)|^2\ S(z_h,t,Q^2,\nu)\ .
\label{160}
 \eeqn
 Here $\alpha$ is the fraction of the quark light-cone momentum carried by
the gluon emitted at the time $t$ by the quark of energy $E_q(t) = \nu -
\Delta E(t)$. The mean radiation time of this gluon $t_c$ is given by
Eq.~(\ref{35}). Assuming that the radiated gluon splits to a $\bar qq$
pair which shares the gluon momentum in equal parts, we conclude that the
leading colorless $q\bar{q}$ pair is produced with an energy
$E_q(t)(1-\alpha/2)$. On the other hand, this must be the observed energy
$z_h\nu$ of the produced hadron, the condition controlled by the first
$\delta$-functions in Eq.~(\ref{160}). The transverse momentum $l_T$ of
the quark and the antiquark within the colorless dipole is related to
$k_T$ as $\vec l_T=3\vec k_T/4$, while the fraction of the hadron
light-cone momentum carried by the antiquark is $\beta(t) =
\alpha(t)/[2-\alpha(t)]$. This is why we need the last two
$\delta$-functions in Eq.~(\ref{160}). The normalization factor $N$ in
(\ref{160}) cancels in ratio (\ref{40}), so we do not specify it.

Note that Eq.~(\ref{160}) has a probabilistic structure; instead of
projecting the production amplitude on the hadron wave function, we
convolute their squares. This simplification is related to our need to
include the time development of hadronization and make use of the
coherence time. To do it correctly one should work with amplitudes taking
care of interferences at all stages. In principle, this can be done within
the path integral approach (partially applied below), however, technically
it is so complicated that the exact solution of this problem does not look
feasible at present. Nevertheless, we will see in what follows that the
approximation Eq.~(\ref{160}) does a good job describing data.

We use the parameterization for the hadronic wave function in
(\ref{160}) in the form of the asymptotic light-cone meson wave function 
\cite{knst},
 \beq
\Psi_h(\beta,l_T^2) \propto
\frac{\beta(1-\beta)}{\beta(1-\beta)+a_{0}}
\exp\biggl [-
\frac{R_h^2~l_{T}^{2}/8}{\beta(1-\beta)+a_{0}}
\biggr],
\label{170}
 \eeq
 where $R_h^2 = 8\langle r_h^2 \rangle_{em}/3$ is related to the mean
hadronic electromagnetic radius squared, $\langle r_h^2 \rangle_{em}$.
Normalization is not important since it can be absorbed into the factor
$N$ in (\ref{160}). The parameter $a_{0}$ in (\ref{170}) incorporates
confinement for highly asymmetric $q\bar{q}$ configurations with
$\beta\rightarrow 0, 1$. We fix this parameter demanding the mean charge
radius of the hadron to be correctly reproduced,
 \beq
\langle \rho^2 \rangle =
\frac{\int \frac{d\beta}{\beta~(1-\beta)}~\int d\rho^2
|\Psi_{h}(\beta,\rho)|^2~\rho^{2} }
{\int \frac{d\beta}{\beta~(1-\beta)}~\int d\rho^2
|\Psi_{h}(\beta,\rho)|^2 } = \frac{8}{3}{\langle r^2 \rangle}_{em}\ ,
\label{180}
 \eeq
 where $\Psi_{h}(\beta,\rho)$ is the hadronic wave function in coordinate
representation, i.e. the Fourier transformation of Eq.~(\ref{170}). For
production of pions with ${\langle r^2 \rangle}_{em} = 0.44 fm^2$
\cite{Amendolia} we found $a_{0}=1/12$.

\begin{figure}[tbh]
\centerline{\psfig{figure=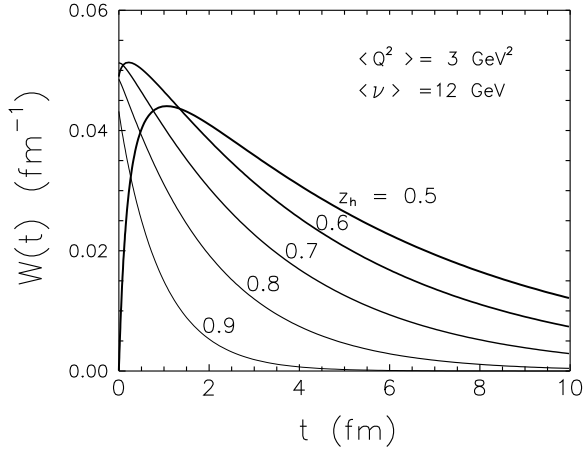,width=7cm}}
\begin{center}\parbox{13cm}{
\protect\caption{Distribution of the hadron production time at
$\nu=12\,$GeV, $Q^2=3\,$GeV$^{2}$ and $z_h=$0.5-0.9}
\label{w} 
}\end{center} 
 \end{figure}

Examples for the production-time distribution $W(t)$ at $<\nu> =
12\,$GeV, $<Q^2> = 3\,$GeV$^2$ and different $z_{h}$ are presented in
Fig.~\ref{w}~\footnote{Note that a function, $t\,W(t)$ was plotted in
Fig.~1 of Ref.~\cite{knp}, but a factor $t$ was missed in the notations.}.
 We see that as a function of $z_h$ the mean production time decreases
similar to the expectations of the string model (\ref{4}) (but with
different value of $\kappa$), which is not a surprise, since this
behavior reflects conservation of energy.

There are no data, of course, for $W(t,z_h,Q^2)$ to compare with, but the
integral over $t$ at fixed $z_h$ is the probability to produce the hadron
with energy $z_h\nu$. This probability is the fragmentation
function which is rather well known phenomenologically:
 \beq
\tilde D_{h/q}(z_{h},Q^2) = \int \limits_0^{\infty} dt\ W(t,z_h,Q^2)\ .
 \label{22}
 \eeq
 Tilde here means that this function describes fragmentation of a quark
only to a (leading) hadron which includes the parent quark.

 Now we can fix the normalization factor $N$ in (\ref{160}). In fact, the
normalization of $\tilde D_{h/q}$ is different from that for the
conventional fragmentation function which is normalized to the particle
multiplicity, $\int_0^1 dz_h\,D_{h/q}(z_{h},Q^2) = \la n_h\ra$.
Apparently, only one hadron in the produced jet contains the original
quark, therefore,
 \beq
\int\limits_0^1 dz_h\,\tilde D_{h/q}(z_{h},Q^2) = 1 .
\label{185}
 \eeq

\begin{figure}[tbh]
\centerline{\psfig{figure=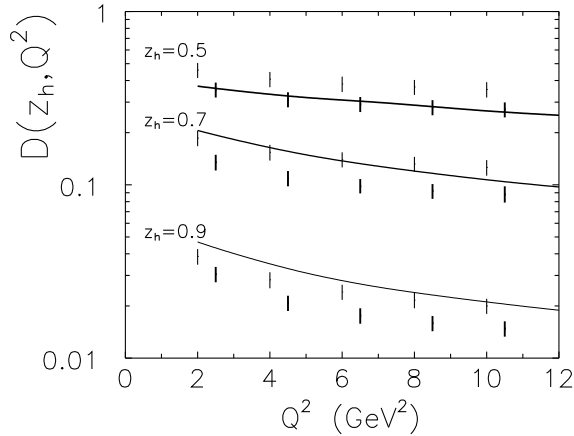,width=7cm}}
\begin{center}\parbox{13cm}{
 \protect\caption{The model prediction for the fragmentation function
$\tilde D(z_h,Q^2)$ in comparison with parametrizations fitted to
data, \cite{bkk} (thin vertical bars) and \cite{kkp} (thick bars).}
 \label{d} 
}\end{center} 
 \end{figure}

Now we are in a position to compare the modeled fragmentation function
with a realistic phenomenological one. In view of the specifics of our
model which is designed to reproduce production of leading hadrons, we
may expect agreement only at large $z_h$, probably $z_h > 0.5$. Our
calculations for $Q^2$ dependence of $\tilde D_{h/q}(z_h,Q^2)$ at
different $z_h$ are depicted in Fig.~\ref{d}, and are compared with
popular parametrizations fitted to data, \cite{bkk} and \cite{kkp} shown
by thin and thick bars respectively. The difference between the
parametrizations may be treated as a systematic uncertainty.
 The model reproduces the phenomenological fragmentation function rather
well. This encourages us to apply this approach to nuclear targets as
well.

\section{Evolution and attenuation of the pre-hadron in the nuclear
environment}

Production of a leading colorless pre-hadron with the detected momentum
completes the process of hadronization. Further propagation of the
pre-hadron in vacuum and formation of the hadronic wave function have no
observable effect, since it leads only to a phase shift. This changes
drastically in the presence of an absorptive medium. In this case the
production time $t_p$ is of special importance. Indeed, when color
neutralization happens and a pre-hadron with the desired energy,
$z_h\nu$, is produced, any subsequent inelastic interaction in the medium
gives rise to a new hadronization process, and the quark energy
substantially degrades down to a lower value of $z_h$. This means that
the pre-hadron should have been produced with a larger $z_h^\prime >
z_h$, i.e. with a smaller probability.  The suppression of the hadron
production rate turns out to be so strong that we can neglect these
double-step processes.

If the energy of the pre-hadron ($\bar qq$ dipole) is sufficiently high
to freeze the transverse $\bar qq$ separation (due to Lorentz time
dilation) for the time of propagation through the medium, then the
survival probability of the pre-hadron, which we also call nuclear
transparency, gets the simple eikonal form \cite{kz},
 \beq
Tr=\left|\frac{\left\la \Psi_h(r_T)\left|
\exp\left[-{1\over2}\sigma_{\bar qq}^N(r_T)\,T_A\right]
\right|\Psi_{\bar qq}(r_T)\right\ra}
{\left\la \Psi_h(r_T)\bigl|\Psi_{\bar qq}(r_T)\right\ra}
\right|^2\ ,
\label{300}
 \eeq
 where $T_A$ is already introduced nuclear thickness, given in this case
by integral of the medium density along the path of the dipole,
$T_A(z_1,z_2) = \int_{z_1}^{z_2}dz\,\rho_A(z)$. The universal
dipole-nucleon cross section, $\sigma_{\bar qq}^N(r_T)$, was introduced
earlier. $\Psi_{\bar qq}(r_T)$ and $\Psi_h(r_T)$ are the light-cone wave
functions of the initial $\bar qq$-dipole (pre-hadron) and the final
hadron respectively.

\begin{figure}[tbh]
 \centerline{\psfig{figure=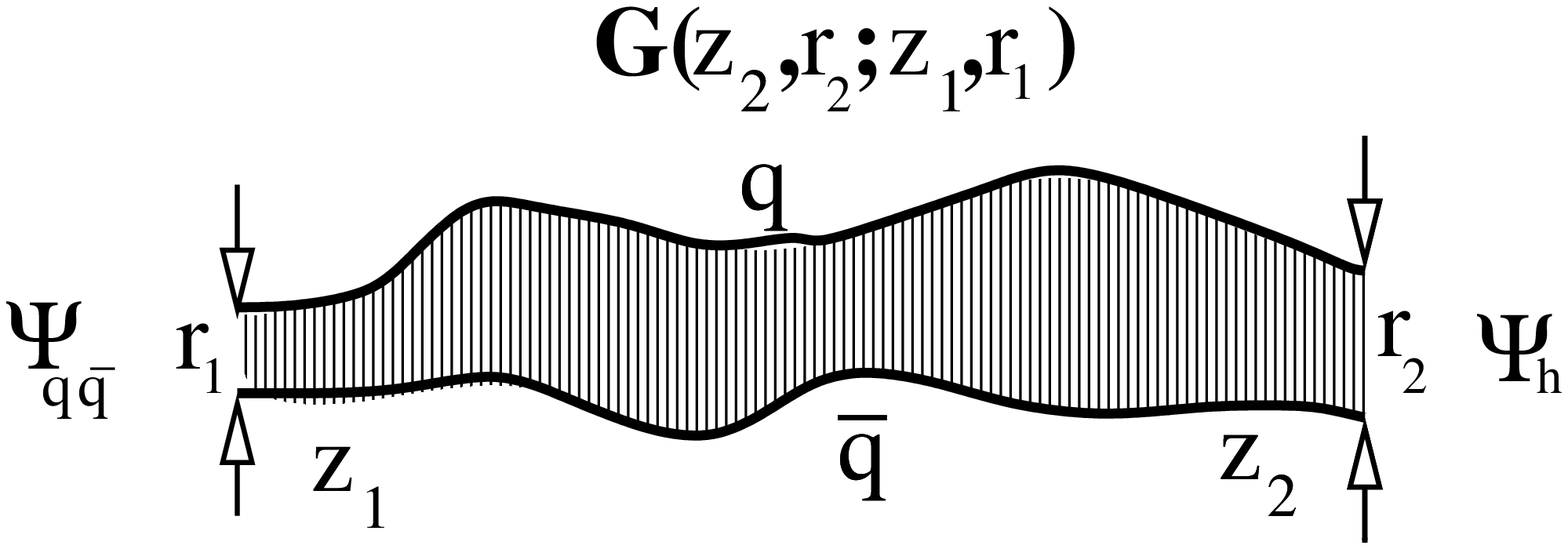,width=8cm}}  
\begin{center}\parbox{13cm}{
 \protect\caption{The light-cone Green function $G_{\bar qq}(z_2,\vec
r_2;z_1,\vec r_1)$ describing propagation of a $\bar qq$ pair over all
possible paths in a medium between points with coordinates $z_1$ and
$z_2$.}
 \label{green} 
}\end{center}
 \end{figure}

In the case of medium energies (HERMES, Jefferson Lab) the dipole size
may substantially fluctuate during propagation through the nucleus and
nuclear transparency might substantially differ from (\ref{300}). The
rigorous quantum-mechanical approach incorporating all effects of
fluctuations should include all possible trajectories of the quark and
antiquark, rather than just fixed quasi-classical ones. This path
integral technique leads to an expression for the survival probability of
a dipole in a medium similar to (\ref{300}). However, the exponential
attenuation factor in the numerator must be replaced by the light-cone
Green function, $G_{\bar qq}(z_2,\vec r_2;z_1,\vec r_1)$, which describes
propagation of the $\bar qq$ pair with initial and final separations
$\vec r_1$ and $\vec r_2$, between points with longitudinal coordinates
$z_{1}$ and $z_{2}$ \cite{kz,kst1,knst}, as is illustrated in
Fig.~\ref{green}.
 Then the nuclear transparency Eq.~(\ref{300}) takes the form,
 \beq
Tr(z_1,z_2)=\left|\frac
{\int d^2r_1 d^2r_2 \Psi^*_h(r_2)
G_{\bar qq}(z_2,\vec r_2;z_1,\vec r_1)
\Psi_{\bar qq}(r_1)}
{\int d^2r\,\Psi^*_h(r)\,
\Psi_{\bar qq}(r)}\right|^2\ .
\label{310}
 \eeq

The Green function satisfies the two-dimensional light-cone Schr\"odinger
equation,
 \beqn
i\frac{d}{dz_2}\,G_{\bar qq}(z_2,\vec r_2;z_1,\vec r_1)&=&
\left[\frac{m_q^{2} - \Delta_{r_{2}}}{2\,E_h\,\beta\,(1-\beta)}
+V_{\bar qq}(z_2,\vec r_2,\beta)\right]\nonumber\\ &\times&
G_{\bar qq}(z_2,\vec r_2;z_1,\vec r_1)\ ,
\label{320}  
 \eeqn
 with the boundary condition $G_{\bar qq}(z_2,\vec r_2;z_1,\vec
r_1)|_{z_2=z_1}= \delta^{(2)}(\vec r_1-\vec r_2)$. Here $E_h=z_h\,\nu$ is
the hadron energy, and $\beta$ is the fraction of the hadron light-cone
momentum carried by the quark. The first, kinetic, term in squared
brackets is responsible for the phase shifts which are related to quark
masses, $m_q$, and their transverse motion described by Laplacian
$\Delta_{r}$ acting on coordinate $r$.

The imaginary part of the light-cone potential in (\ref{320}),
 \beq
{\rm Im}\,V_{\bar qq}(z,\vec r,\beta)=
-{1\over2}\,\sigma^N_{\bar qq}(r)\,\rho_A(z)\ ,
\label{330}
 \eeq
 is responsible for attenuation of the $\bar qq$ in the medium, while the
real part represents the interaction between the $q$ and $\bar{q}$.  The
latter is supposed to provide the correct light-cone wave function of the
final meson. For the sake of simplicity we use the oscillator form of the
potential,
 \beq
{\rm Re}\,V_{\bar qq}(z,\vec r,\beta) =
\frac{a^4(\beta)\,\vec r\,^2} 
{2\,E_h\,\beta(1-\beta)}\ ,
\label{340} 
 \eeq
 which leads to a Gaussian $r$-dependence of the light-cone wave function
of the meson ground state.  The shape of the function $a(\beta)$ was
discussed in \cite{kst2}, and fixed by photoproduction data
$a^2(\beta)=a^2_0+a^2_1\beta(1-\beta)$, where
$a_0^2=v^{1.15}\times(112\MeV)^2$;  
$a_1^2=(1-v)^{1.15}\times(165\MeV)^2$, and $v$ takes any value in the
interval $0<v<1$.

 If we assume also that the small-$r$ dependence for the dipole cross
section, $\sigma^N_{\bar qq}(r)= C(E_h)\,r^2$, then Eq.~(\ref{320}) has an
analytic solution \cite{fg},
 \beqn 
&& G_{\bar qq}(z_2,\vec r_2;z_1,\vec r_1) =
\frac{a^2(\beta)}{2\;\pi\;i\;
{\rm sin}(\xi\,\Delta z)}
\nonumber\\ &\times&
{\rm exp}
\left\{\frac{i\,a^2(\beta)}{2{\rm sin}(\xi\,\Delta z)}\,
\Bigl[(r_1^2+r_2^2)\,{\rm cos}(\xi \;\Delta z) -
2\;\vec r_1\cdot\vec r_2\Bigr]\right\}
\nonumber\\ &\times& 
{\rm exp}\left[- 
\frac{i\,m_q^{2}\,\Delta z}
{2\,E_h\,\beta\,(1-\beta)}\right] \ , 
\label{350} 
 \eeqn
where $\Delta z=z_2-z_1$ and 
 \beq 
\xi = \frac{\sqrt{a^4(\beta)-i\beta(1-\beta)E_h\,C(E_h)\,\rho_A}}
{E_h\;\beta(1-\beta)}\ ,
\label{360} 
 \eeq
 This solution is valid only for a medium with a constant density.
However, using the recurrent relations between Green functions for
different space intervals one can analytically solve the problem for a
varying density as well \cite{kz}.

\section{Nuclear attenuation caused by absorption}

We assume that the colorless $\bar qq$ pair (pre-hadron) produced at
distance $l_p$ from the DIS point has a Gaussian distribution of
transverse separations with the mean transverse size related to the
inverse value of the mean transverse momentum of the quarks,
 \beq
\langle r_1^2 \rangle = 
\frac{4}{\langle l_T^2(t_p) \rangle} =
\frac{64}{9~\langle k_T^2(t_p) \rangle}
\label{370}
 \eeq
 where
 \beq 
\langle k_T^2(t) \rangle = 
\frac{\int_{\lambda^2}^{Q^2}
dk_{T}^2\,\Bigl[d{W}(t,z_h,k_T^2,\nu)/d\ln(k_T^2)\Bigr]} 
{\int_{\lambda^2}^{Q^2}
dk_{T}^{2}\,\Bigl[d{W}(t,z_h,k_T^2,\nu)/dk_T^2\Bigr] }\ .
\label{380}
 \eeq

A $\bar qq$ dipole with starting value of transverse size $r_1(t_p)$
experiences evolution, forming the hadronic wave function and getting
through the filtering process in the nuclear medium which absorbs dipoles
of larger size more effectively. If it were in vacuum, it would take
formation time $t_f$ which can be estimated as
 \beq
t_f \approx \frac{2z_{h}\nu}{m^2_{h'}-m^2_h}\ ,
\label{390}
 \eeq
 where $h'$ is the first radial excitation of the hadron $h$.  As was
mentioned in the introduction, the formation time linearly rises with
$z_h$, while the production time vanishes at $z_h \to 1$.

The effective fragmentation function modified by the survival probability
(transparency) of the pre-hadron in the nuclear matter reads,
 \beqn
&&D^{A}_{eff}(z_{h},Q^2,\nu)	=
\int d^2{\bf b} \int \limits_{-\infty}^{\infty} dz\,
\rho_{A}({\bf b},z)\nonumber\\&\times&
\int_{0}^{\infty} dt\, W(t,z_h,Q^2,\nu)\,
Tr(b,z+t,\infty)\ .
\label{400} 
 \eeqn
 In the case of vanishing absorption, $\sigma_{\bar qq}^N\to0$, this
fragmentation function is identical to that of the free nucleon, except it
is $A$ times larger. The ratio of the effective fragmentation function on
a nuclear target to one on a free nucleon is identical to the experimental
definition Eq.~(\ref{40}).

As was mentioned, we use the dipole approximation, $\sigma_{\bar qq}^N(r)=
C\,r^2$. The factor $C$ is proportional to the inverse mean hadron radius
squared which is related to the parameters of the oscillatory potential,
 \beq
C = \frac{\sigma^{hN}_{tot}}{\la r_T^2\ra_{h}}=
\frac{1}{2}\,m_{q}\,\omega\,\sigma_{tot}^{\pi N}\, .
\label{410}
 \eeq
 where $\omega = (m_{h'}-m_{h})/2$ is the oscillatory frequency. For
production of leading pions we took $\omega=0.3$\,GeV and
$m_{q}=0.2$\,GeV which is constrained by the mean charge pion radius,
 \beq
\la r_T^{2}\ra_h = \frac{2}{m_{q}\,\omega} =
\frac{8}{3}\,<r^{2}>_{ch}\ ,
\label{420}
 \eeq
 and nuclear shadowing data \cite{krt}. Using $\sigma_{tot}^{\pi N}=
25\mb$ we arrive at $C\sim 1.9$.

Note that the dipole cross section must be flavor independent, i.e. factor
$C$ must be the same for different species of mesons. This condition is
satisfied by Eq.~(\ref{410}) as long as $\sigma^{hN}_{tot} \propto
\la r_T^2\ra_h$. Such a relation between radiuses of hadrons and their cross
sections is supported by data \cite{hp}.

\section{Comparison with data}

Unfortunately the low statistics of the HERMES data does not allow to
perform a multi-dimensional comparison with the theory. One should
integrate over all variables except one to get data with reasonable
errors.

\subsection{Energy dependence of nuclear effects}

\begin{figure}[tbh]
 \centerline{\psfig{figure=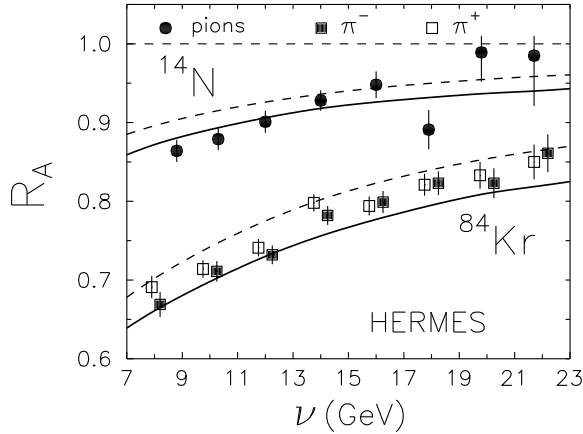,width=7cm}}  
\begin{center}\parbox{13cm}{
 \protect\caption{Comparison of the model predictions for the energy
dependence of the nuclear ratio Eq.~ (\ref{40}) for nitrogen and krypton
targets HERMES data \cite{hermes}. The solid and dashed curves correspond   
to calculations with and without the effects of induced radiation
respectively}
 \label{pi-nu} 
}\end{center} 
 \end{figure}

The results of numerical calculations for energy dependence of the ratio
$R_A(\nu)$ defined in (\ref{40})
are depicted by dashed curve in
Fig.~\ref{pi-nu} in comparison with with HERMES data. 
 The cross sections for leading pion production were integrated over
$z_h$ and $p_T$ at the mean value of $Q^2$ same as in HERMES data. In
what follows we always fix implicit variables at the values
corresponding to the kinematics of HERMES\footnote{We are grateful to
Valeria Muccifora and Pasquale Di Nezza for providing us with required
experimental information.}.

Note that we ignored the induced energy loss in these calculations and
integrated over $z_h > 0.5$. Since the data are integrated over
$z_h>0.2$, this may cause some difference. We also neglected nuclear
effects in data taken on a deuteron target when it is used as the
denominator in (\ref{40}).

It is clear why nuclear suppression is maximal at low energies and has a
tendency to vanish with increasing energy (data from EMC experiment
\cite{emc} at much higher energies indeed demonstrate no nuclear
effects). As we discussed in previous sections, hadronization process is
completed by production of the leading colorless wave packet, which is
then evolving into the final hadron over the formation time $t_{f}$ given
by (\ref{390}). If color neutralization happens outside the nucleus
($t_p>R_A$), the evolution leads only to a phase shift, resulting in no
nuclear effects, as long as we neglect the induced energy loss.

On the other hand, the evolution may cause a strong attenuation if the
wave packet is produced inside the nuclear target, as it happens at low
energies where $t_p< R_{A}$. At higher energies $t_p$ rises leaving less
room for absorption. Besides, the formation time $t_f$ rises too, and the
initial small size, $r_T\sim 1/Q$, of the pre-hadron is evolving slower,
leading to a lesser attenuation due to color transparency. Both effects
lead to a stronger nuclear suppression at small than at high energies in
accordance with data from HERMES \cite{hermes} (Fig.~\ref{pi-nu}), SLAC
\cite{slac} and EMC \cite{emc,pavel} experiments.

\subsection{\boldmath$z_h$-dependence of nuclear effects}

\begin{figure}[tbh]
 \centerline{\psfig{figure=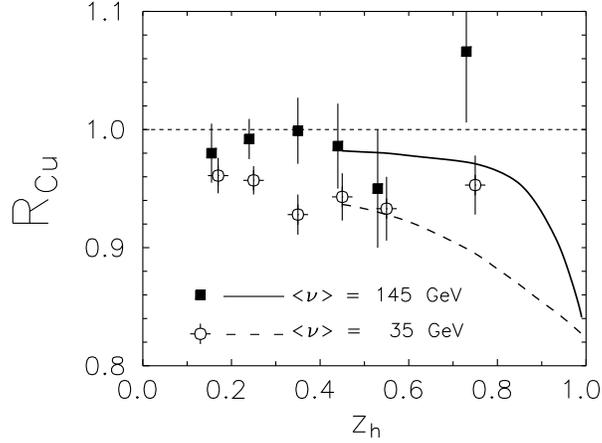,width=7cm}}  
\begin{center}\parbox{13cm}{
 \protect\caption{Comparison of the model calculations for the
$z_h$- dependence of nuclear suppression $R_{Cu}$ (\ref{40}) at the
energy $<\nu>=35$\,GeV (dashed curve) and $<\nu>=145$\,GeV (solid
curve) with the EMC data \cite{emc,pavel}. Calculations are done at
$Q^2=11\GeV^2$.}
 \label{emc} 
}\end{center} 
 \end{figure}

Even at high energies the production time shrinks at large $z_h\to 1$
according to (\ref{4}) and Fig.~\ref{w}. Thus, at large $z_h$ the
path available for absorption of the pre-hadron reaches maximal length
resulting in a strongest nuclear suppression. Such a shape of
nuclear ratio falling toward $z_h=1$ has been expected since 30 years
ago \cite{kn,bg}, but high energy experiments could not reach
sufficiently large values of $z_h$. Our predictions \cite{knp} are
compared with EMC data in Fig.~\ref{emc}.
 We see that the interval of $z_h$ where $l_p$ is short absorption is at
work, is rather short and squeezed towards $z_h=1$. This is because the
energy of the EMC experiment is too high, but absorption is at work only
if $l_p\propto \nu(1-z_h)$ is shorter than the nuclear size, what needs
quite a small $(1-z_h)\propto 1/\nu$.

\begin{figure}[b]
 \centerline{\psfig{figure=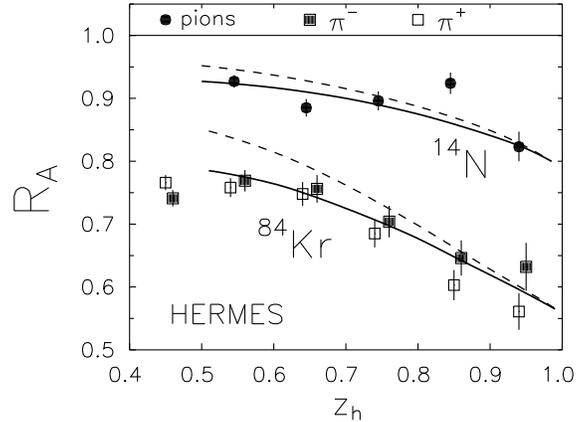,width=7cm}}  
\begin{center}\parbox{13cm}{
 \protect\caption{Model predictions for the $z_{h}$- dependence of
nuclear ratios $R_{^{14}N}$ (\ref{40}) and $R_{^{84}Kr}$ calculated at
values of $<\nu>$ and $Q^2$ corresponding to the measurements at HERMES.
The solid and dashed curve are the model predictions with and without
inclusion of induced radiation, respectively.}
 \label{pi-zh} 
}\end{center} 
 \end{figure}

This is why measurements at much lower energies of HERMES were proposed
in \cite{knp} in order to detect this effect. Fig.~\ref{pi-zh}
demonstrates HERMES data for nitrogen and krypton in comparison with our
predictions shown by dashed curves for the case of pure absorption.
 This is the first experimental confirmation for the longstanding
prediction of vanishing production length at $z_h\to 1$.

Note, however, that even at $z_h\to1$ nuclear suppression vanishes if
both $\nu$ and $Q^2$ are sufficiently large. This is demanded in QCD by
$k_T$-factorization and is correctly reproduced by our model. Indeed, at
high $Q^2$ the initial size of the pre-hadron is very small, and it
remains small long enough if the energy is high. Then color transparency
eliminates nuclear effects.

\subsection{\boldmath$Q^2$ dependence of nuclear effects}

\begin{figure}[b]
 \centerline{\psfig{figure=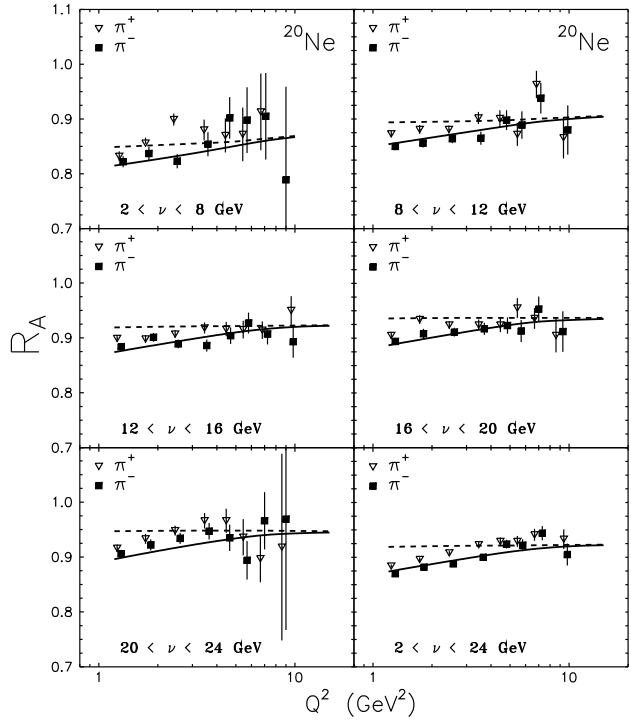,width=8cm}}  
\begin{center}\parbox{13cm}{
 \protect\caption{Comparison of the model calculations for the
$Q^{2}$- dependence of the ratio $R_{Ne}$ of cross sections 
integrated over $z_h$ with preliminary data from the HERMES experiment 
\cite{hermes-q2}. The solid and dashed curve are the model
predictions with and without corrections for induced gluon radiation 
respectively.}
 \label{q2-ne} 
}\end{center} 
 \end{figure}

At least two effects are sensitive to a variation of $Q^2$, color
transparency and the production length. Apparently, the initial size of
the pre-hadron shrinks at larger $Q^2$ and the nuclear matter becomes more
transparent, i.e. the survival probability for the pre-hadron increases.
At the same time, the production length $l_p$ contracts because of rising
vacuum energy loss, and provides a longer path available for absorption.
This effect leads to a stronger nuclear suppression at larger $Q^2$. Thus,
the two effects work in opposite directions and may partially or fully
compensate each other. Indeed, it nearly happens: our predictions for the
effect due purely to absorption depicted by dashed curves in
Fig.~\ref{q2-ne} for neon demonstrate a rather weak $Q^2$-dependence.
 Although the $Q^2$-dependence demonstrated by the HERMES data is not
steep either, one can see some difference. We will come to this point
later after inclusion of corrections related to induced energy
loss.

\section{Corrections for induced gluon radiation}

Multiple interactions of the quark during the time interval between the
DIS point and production of the colorless pre-hadron cannot stop or absorb
the quark, which keeps propagating through the medium, rotating in the
color space. Nevertheless, the additional soft kicks gained by the quark
force it to shake off more gluons. This additional gluon radiation induced
by the medium increases the loss of energy compared to vacuum. We
calculate the induced energy loss in accordance with Eq.~(\ref{30}). This
result of \cite{bdmps} has corrections for finite medium and high energies
\cite{glv,zakharov}. However, the energies of HERMES are not high and the
nuclear size is much longer than the gluon mean free path which is of the
order of $0.2-0.3\fm$ \cite{pisa,k3p,sz}.

$\Delta p_T^2$ in Eq.~(\ref{30}) can be taken from data on
$p_T$-broadening in Drell-Yan reaction, or from calculations (see next
section) which agree with data. Then one should add this induced energy
loss to the vacuum one, i.e. to replace $\Delta E_{vac}(t)\Rightarrow
\Delta E_{vac}(t) + \Delta E_{ind}(t)$ for the time-dependent quark
energy, $E_q(t)\equiv \nu -\Delta E(t)$ in Eq.~(\ref{160}).

Another modification of our calculations caused by $p_T$-broadening in
the medium is the modification of the hard scale. Namely, one should
replace $Q^2 \Rightarrow Q^2+\Delta p_T^2(t)$ in Eq.~(\ref{150}) for the
Sudakov factor and in Eq.~(\ref{160}).

Both corrections enhance nuclear suppression, i.e. diminish the nuclear
ratio. The results are shown by solid curves in Figs.~\ref{pi-nu},
\ref{pi-zh}, \ref{q2-ne} and \ref{q2-kr}. Apparently, these induced energy
corrections vanish if $l_p \to 0$. This happens at $z_h \to 1$ and at
maximal $Q^2$. Indeed, the dashed and solid curves in Figs.~\ref{pi-zh},
\ref{q2-ne} and \ref{q2-kr} coincide in these limits.

Note that inclusion of induced energy loss improves agreement with data.
This fact may be considered as an indication to the importance of these     
corrections, although one should not probably expect very good description
of data by such a rough model.

\section{Nuclear broadening of the transverse momentum distribution}

According to the perturbative QCD treatment of induced energy loss, it is
caused by the broadening of transverse momentum of the quark initiating a
jet during the hadronization process (at $t<t_p$).  At longer time
intervals, $t>t_p$, the produced pre-hadron carries undisturbed
information about the quark transverse momentum (since it has no
interactions) which thus can be directly measured. Comparison with
theoretical predictions seems to be of special importance.

A parton propagating through a medium experiences multiple interactions
in the medium and performs Brownian motion in transverse momentum plane.
As a result, the mean transverse momentum squared of the quark linearly
rises with the path length. Quantitatively this process has been
successfully described within the light-cone dipole approach
\cite{dhk,jkt} (otherwise, most of models just fit the data to be
explained). If the DIS takes place on a nucleon with coordinates $(\vec
b,z)$, the medium modified transverse momentum distribution is expressed
in terms of the universal cross section of a $\bar qq$ dipole with a
nucleon,
 \beqn
&&\frac{dN_q^A(b,z)}{d^2p_T} = \int d^2r_T\,d^2r^\prime_T\,
\Omega_q(\vec r_T,\vec r_T^\prime)\,
\exp\left[i\vec p_T(\vec r_T-\vec r_T^\prime)\right]
\nonumber\\
&\times&
\exp\left[-\frac{T_A(b,z,z+l_p)}{2}\,
\sigma_{\bar qq}^N(\vec r_T-\vec r_T^\prime,\nu)\right].
\label{1000}
 \eeqn
 Here $T_A(b,z,z+l_p)=\int_z^{z+l_p}dz'\,\rho_A(b,z')$ is the nuclear
thickness function; $\sigma_{\bar qq}^N(r_T,\nu)$ is the phenomenological
dipole cross section fitted to data for the proton structure function and
photoabsorption cross section. It includes nonperturbative effects,
as well as all corrections for gluon radiation (which give rise to the
energy dependence of the cross section) also contributing to
the broadening of the quark transverse momentum. 

The quark density matrix in coordinate representation,
 \beq
\Omega_q(\vec r_T,\vec r_T^\prime) =
\frac{k_0^2}{\pi}\,
\exp\left[-{k_0^2\over2}(r_T^2+r_T^{\prime\,2})\right]\ ,
\label{1100}
 \eeq
 where $k_0$ is the mean transverse momentum of the quarks, describes the
initial distribution of the quark in the target nucleon. It controls the
transverse momentum distribution of the valence quark emerging from DIS on
a free nucleon. Therefore, the mean transverse momentum is of the order of
the inverse proton size, i.e. is rather small, $k_0\sim\Lambda_{QCD}$.
Usually data on hard processes need a much larger primordial transverse
momentum. This effect is related to the next-to-leading order corrections
related to hard gluon emission. Since we explicitly include gluon
radiation in our model, additional next-to-leading order corrections would
lead to double counting.

Note that our parameter-free approach successfully explained available
data for the Cronin effect in $pA$ collisions in fixed target experiments,
and correctly predicted it for $d-Au$ collisions at RHIC
\cite{prl,phenix}. Data for nuclear broadening in Drell-Yan reaction also
are well reproduced \cite{jkt}.

To get the hadron transverse momentum distribution we should convolute
the quark $p_T$-distribution, Eq.~(\ref{1000}) calculated without the
exponential factor, with the fragmentation function $D_{h/q}(z_h,p_T)$,
 \beqn
&& \frac{1}{\sigma^N_{DIS}}\,
\frac{d\sigma^N_{DIS}(\gamma^*N\to hX)}{d^2p_T} =
\int d^2 k_T\,d^2q_T\,\frac{dN_q^N}{d^2q_T}
\nonumber\\ &\times&
\frac{dD_{h/q}(z_h,k^2_T)}{d^2k_T}\,
\delta\left(\vec p_T-\vec q_T-
\frac{\vec k_T}{2}\right)\ .
\label{1200}
 \eeqn

Now we can make use of our model for the vacuum fragmentation function
$\tilde D_{h/q}(z_h,Q^2)$, Eq.~(\ref{185}), and calculate the mean
momentum squared of the produced meson. It hardly varies with $Q^2$
within the kinematic range corresponding to HERMES data. We found $\la
p_T^2\ra=\la k_T^2\ra/4+k_0^2=0.3\GeV^2$, where $k_0=0.2\GeV$. This
result agrees with the $p_T$-distribution of hadrons measured at SLAC
\cite{slac-pt} at energies overlapping with the HERMES energy range.

The $p_T$-dependence of the effective fragmentation function in the
nuclear medium, according to Eq.~(\ref{400}), reads,
 \beqn
&& \frac{A}{\sigma^A_{DIS}}\,
\frac{d\sigma^A_{DIS}(\gamma^*A\to hX)}{d^2p_T} =
\int d^2{\bf b} \int \limits_{-\infty}^{\infty} dz\,
\rho_{A}({\bf b},z)
\nonumber\\&\times&
\int_{0}^{\infty} dt\ Tr(b,z+t,\infty)
\int d^2k_T\,d^2q_T\,\frac{dN^A_q(b,z)}{d^2q_T}\,
\nonumber\\&\times&
\frac{dW(t,z_h,k^2_T,\nu)}{d^2k_T}\,
\delta\left(\vec p_T-\vec q_T-
\frac{\vec k_T}{2}\right)
\label{1300} 
 \eeqn
 We also included the small effect of Fermi motion in the nucleus.
 
\begin{figure}[tbh]
 \centerline{\psfig{figure=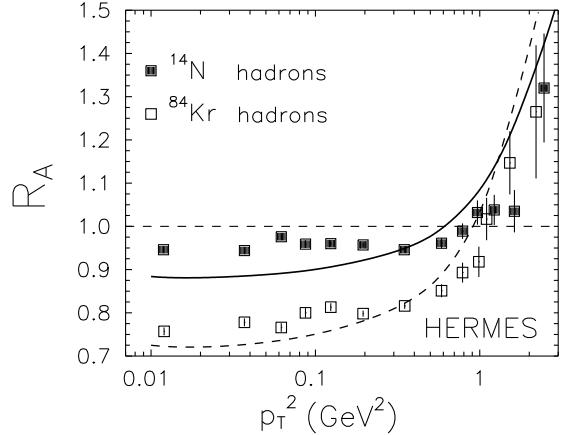,width=7cm}}  
\begin{center}\parbox{13cm}{
 \protect\caption{The $p_T$-dependent nucleus-to-nucleon ratios of the
cross sections of hadron production in DIS on nitrogen (solid curve) and
krypton (dashed curve) in comparison with HERMES data.}
 \label{pt} 
}\end{center} 
 \end{figure}

The calculated ratio of the two distributions, nucleus-to-nucleon, is
depicted in Fig.~\ref{pt} for nitrogen and krypton in comparison with
data from the HERMES experiment \cite{hermes}.
 The model reproduces well the data with no adjustment, although the
curves seem to rise somewhat steeper than the data at large $p_T$. That
may be caused by the difference in the bottom limits of integration over
$z_h$ ($0.2$ in the data, versus $0.5$ in our calculations).

\section{Disentangling absorption and induced energy loss}

The production length is a fundamental characteristics of the early stage
of hadronization, and it is important to sort out those observables
which are sensitive to $l_p$.

Such observables could also help to disentangle different models of
hadronization. The current estimates of $l_p$ within different models
(strings \cite{kn,bg,hermes,amp}, pQCD \cite{knp}) are pretty close and
show that $l_p$ is rather short compared to the nuclear sizes for the
kinematics of the experiments at HERMES \cite{hermes} and especially at
Jefferson Lab \cite{will}. At the same time, some models \cite{ww} make
an assumption that color neutralization happens always outside the
nucleus. Although this is an ad hoc assumption with no justification,
would be useful to confront it with direct experimental evidence.

\subsection{Flavor dependence}

The survival probability for the produced pre-hadron depends on the wave
function of the final hadron, as one can see in Eq.~(\ref{310}). Also
intuition tells us that the projection to a wave function with a larger
mean radius should emphasizes larger $\bar qq$ separations in the
pre-hadron, leading to a stronger absorption. Since the pion has a larger
radius than the kaon, the former is expected to be produced with a
stronger nuclear suppression than the latter. This may be a good
signature for absorptive effects and shortness of $l_p$. On the contrary,
in the case of $l_p\gg R_A$ there should be no sensitivity to the
hadronic size.
 \begin{figure}[tbh]
 \centerline{\psfig{figure=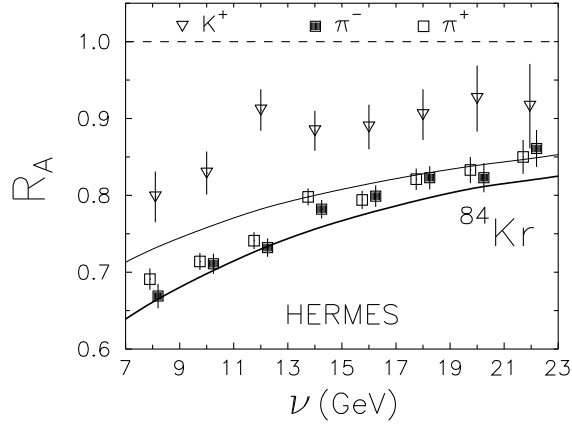,width=7cm}}  
\begin{center}\parbox{13cm}{
 \protect\caption{Comparison of nuclear effects for kaon (thin curve) and
pion (thick curve) productions as function of energy in comparison with
HERMES data.}
 \label{k-pi-nu} 
}\end{center} 
 \end{figure}
The two curves are different only by the value of the charge radius, $\la
r^2\ra_{ch}$ (kaon or pion) used in calculations.

The HERMES experiment has recently provided high quality data for hadron
production with particle identification \cite{hermes}. Here we concentrate
only on production of pions and positive kaons, since negative kaons and
antiprotons do not contain valence quarks common to the target nucleon,
while proton production may be contaminated with the target nucleons.

Our predictions for the production rates of positive kaons and pions
integrated over $z_h$ are depicted in Fig.~\ref{k-pi-nu} as function of
photon energy together with recent data from HERMES. 

 Data confirm the expected identical suppression for positive and
negative pions, and less attenuation for kaons. Although our calculations
underestimate kaons, this is due to the difference in the bottom limit of
integration over $z_h$, $(z_h)_{min}=0.2$ in the data, and $0.5$ in our
calculations. At small $z_h$ kaons are copiously produced via reaction
$\pi p\to K\Lambda$ which has a very low energy threshold. It would make
sense if experimental data used a larger value of $(z_h)_{min}$.
 \begin{figure}[tbh]
 \centerline{\psfig{figure=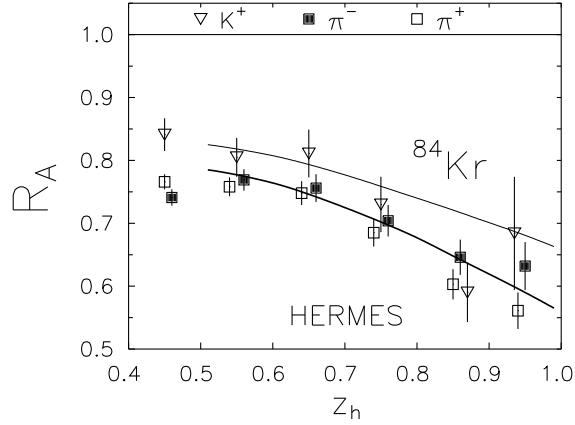,width=7cm}}  
\begin{center}\parbox{13cm}{
 \protect\caption{Nuclear ratios for pions (thick curve) and positive
kaons (thin curve) as function of $z_h$ predicted by our model in
comparison with HERMES data.}
 \label{k-pi-zh} 
}\end{center} 
 \end{figure}

Although the alternative model, the energy loss scenario, is not sensitive
to the hadronic radius, one may expect a difference in quark fragmentation
functions for $q\to \pi$ and $q\to K$, where $q$ is a light quark, $u$ or
$d$. Indeed, the end-point behavior of the fragmentation function, at
$z_h\to 1$, is dictated by the Regge phenomenology \cite{dpm},
 \beqn
D_{\pi/q}(z_h) &\propto& (1-z_h)^{1-\alpha_{\rho}(0)} =
(1-z_h)^{0.5}\ ;
\nonumber\\
D_{K^+/q}(z_h) &\propto& (1-z_h)^{1-\alpha_{K^*}(0)} =
(1-z_h)^{0.8}\ .
\label{pi-k}
 \eeqn
 Since the induced energy loss leads to a shift in $z_h \Rightarrow z_h
+\Delta z_h$, it should lead to a stronger suppression 
of kaons which have a steeper fragmentation function.
This expectation is at variance with HERMES data
depicted in Fig.~\ref{k-pi-zh}.

 At the same time, the data agree well with our predictions shown by
thick and thin curves for pions and positive kaons respectively. This
fact confirms our claim that the disagreement with data for kaons seen in
Fig.~\ref{k-pi-nu} is due to presence of small-$z_h$ events in the data.

Note that the maximum effect of flavor dependence should be expected for
shortest $l_p$, e.g. at $z_h\to1$. Indeed, our calculations predict quite
a large difference between nuclear effects for $K^+$ and pions at
$z_h\to1$, as is demonstrated in Fig.~\ref{k-pi-zh}. Unfortunately, the
accuracy of the data is not sufficient to see such a variation of the
flavor dependence with $z_h$.

\subsection{\boldmath$Q^2$-dependence}

The fit to HERMES data performed in Ref.~\cite{ww} within the energy loss
scenario led to the value of the universal parameter, twist-4 quark-gluon
correlation tensor, which is twice as small as extracted from Drell-Yan
data and $25$ times smaller than follows from data on high-$p_T$ dijet
production. Such a steep variation of the parameter which must be
process-independent was attributed in \cite{ww} to a strong scale
dependence. If it were true, the induced energy loss effect would rise
with $Q^2$. Accordingly, the nuclear suppression should become stronger
at larger $Q^2$. This expectation is in obvious contradiction with the
HERMES results depicted in Fig.~\ref{q2-ne}.
\begin{figure}[bht]
 \centerline{\psfig{figure=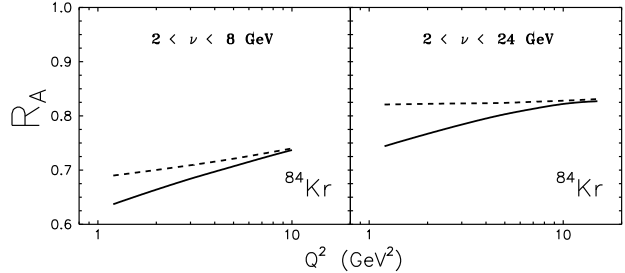,width=8cm}}  
\begin{center}\parbox{13cm}{
\protect\caption{The same as in Fig.\ref{q2-ne}, but predicted for
krypton.}
 \label{q2-kr} 
}\end{center} 
 \end{figure}

On the other hand, one may expect a more pronounced rise of the nuclear
ratio with $Q^2$ for heavy nuclei. Our results for krypton depicted in
Fig.~\ref{q2-kr} confirm this expectation, although the difference from
data on neon target, Fig.~\ref{q2-ne} is not a dramatic effect.
Note the sizeable variation of nuclear suppression with $Q^2$ at low
energies. In this case the production time is short at any $Q^2$ and the
variation of nuclear effects versus $Q^2$ is only due to color
transparency. This is why the nuclear ratio rises with $Q^2$ even without
induced energy loss corrections.

\subsection{\boldmath$z_h$ dependence of \boldmath$p_T$-broadening}

\begin{figure}[tbh]
 \centerline{\psfig{figure=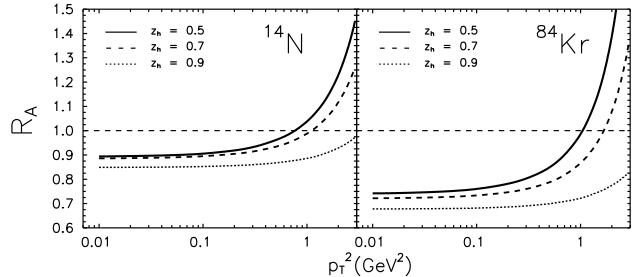,width=8cm}}  
\begin{center}\parbox{13cm}{
 \protect\caption{$p_T$-dependence of nucleus-to-nucleon ratios 
binned in $z_h$ for hadroproduction in DIS on nitrogen (left) and
krypton (right).}
 \label{pt-zh} 
}\end{center} 
 \end{figure}

According to the discussion in the previous section, nuclear broadening of
the transverse momentum distribution might be the most sensitive probe for
the production length, and provides a direct measurement of $l_p$, since
$\Delta p_T^2 \propto l_p$. Indeed, only the hadronizing quark at short
time intervals $t<t_p$ contributes to the broadening. As soon as the
pre-hadron is created, no further broadening occurs, since inelastic
interactions are prohibited for the pre-hadron, and only broadening via
elastic rescattering is possible. However, the elastic cross section is
so small that even for pions the mean free path in nuclear matter is
about $20\fm$. It is even longer for a small-size pre-hadron due to color
transparency. Therefore, we should expect a disappearance of the
broadening effect at large $z_h\to 1$ since $t_p\to 0$. In
Fig.~\ref{pt-zh} we show our predictions for the nuclear ratio as
function of $p_T$ for different $z_h$ bins.

 Note, however, that broadening should be weaker at larger $z_h$ for any
model which is able to describe the observed $z_h$-dependence of the
nuclear ratio (see Fig.~\ref{pi-zh}). Indeed, the fact that $R_A(z_h)<1$
implies that DIS happens only in a part of the nuclear volume, namely
its back side. Then, the fact that $R_A(z_h)$ decreases towards $z_h=1$
means that a smaller part of the nuclear volume is working at larger
$z_h$, i.e. the DIS reaction is pushed towards the back surface of the
nucleus. Therefore, even in the energy loss model the path available for
broadening (from the DIS point and on) becomes shorter at large $z_h$.
This should also lead to a reduction of the broadening, but rather small,
only about $10\%$ for nitrogen and $20\%$ for krypton. At the same time,
in our approach the broadening completely vanishes at $z_h\to1$ since
$l_p\to 0$. It would be extremely important for HERMES and the experiment
running at Jefferson Lab \cite{will} to provide relevant data.

\subsection{Double hadron production}

One may think that a process of double hadron production is a sensitive
tool which is able to disentangle the inside-outside hadronization. If
absorption of the pre-hadrons is the main contributor to nuclear
suppression, naively one would expect the nuclear ratio for double-hadron
production to be squared, i.e. quite small, compared to inclusive
single-hadron channel.

A closer look, however, shows that the situation is quite complicated and
such an expectation is oversimplified. Let us assume that the two
pre-hadrons are created inside the nuclear medium with production lengths
$l_1$ and $l_2$ respectively, and $l_2 > l_1$. For the sake of
simplicity, just for this consideration, we assume that the nuclear
density is constant, both hadrons have the same constant absorption
cross sections $\sigma$, and we neglect any additional suppression
related to induced energy loss. Then it is straightforward to write the
nuclear suppression ratio at given impact parameter $b$,
 \beqn
R^{(2h)}_A(b) &=&
\int\limits_{-\infty}^\infty
dz\,\rho_A(b,z)\,
\exp\Bigl[-\sigma\int\limits_{z+l_1}^\infty
dz'\,\rho_A(b,z')\Bigr]\ 
\nonumber\\ &\times&
\exp\Bigl[-\sigma\int\limits_{z+l_2}^\infty
dz'\,\rho_A(b,z')\Bigr]
\label{2050}
 \eeqn
 Using the approximation of a constant nuclear density,
$\rho_A(r)=\rho_A\Theta(R_A-r)$, and neglecting $\exp(-2\sigma\rho_A L)\ll
1$, where $L=\sqrt{R_A^2-b^2}$, we arrive at the relative, double to
single hadron production, nuclear suppressions,
 \beq
\frac{R^{(2h)}_A(b)}{R^{(1h)}_A(b)}=
\frac{1-
{1\over2}\exp(-\sigma\rho_A\Delta l)+
\sigma\rho_A l_1}
{1+\sigma\rho_A l_p}\ ,
\label{2060}
 \eeq
 where $\Delta l = l_2-l_1$ and $l_p$ is the single-hadron production
length.  This ratio varies dependent on the values of $l_1,\ l_2,\ l_p$
and the mean free path, $1/\sigma\rho_A$, in the medium. In the case of a
very dense matter, e.g. produced in heavy ion collisions, one should
expect $l\sigma\rho_A\gg1$. Contrary to naive expectations, the
production rates of single and double hadrons according to (\ref{2060})
should be suppressed equally.  The data for high-$p_T$ production at RHIC
(see next section) seem to confirm this. This situation may change at
very large $p_T$ due to contraction of the production lengths [see
Eq.(\ref{2030})].

In the case of DIS in the dilute nuclear medium production lengths are
comparable with the mean free path, which is about $2\fm$. Then the
result should vary dependent on values of $z_h$ for each of the hadrons.
For instance, if values of $z_h$ in both cases approach their maximal
values, i.e.  $z_h\to 1$ and $z_{h1}+z_{h2}\to 1$, all production lengths
vanish, and the ratio Eq.~(\ref{2060}) equals $1/2$, i.e. a double-hadron
is suppressed twice as much as a single-hadron.

\section{Heavy ion collisions: Energy loss or absorption?}

\begin{figure}[tbh]
 \centerline{\psfig{figure=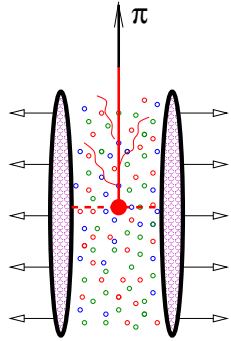,width=3cm}}  
\begin{center}\parbox{13cm}{
 \protect\caption{Propagation and hadronization of a high-$k_T$ parton
in the dense partonic matter created right after collision of two nuclei
in their c.m. frame.}
 \label{aa} 
}\end{center}
 \end{figure}

The situation in the process of high-$p_T$ hadron production in 
heavy ion collisions at very high energies is similar to the process of 
hadron production in DIS off nuclei at medium energies. 
Indeed the cartoon in Fig.~\ref{aa} shows the development of
hadronization after the nuclear disks passed through each other.
 Partons are copiously created from vacuum forming a dense medium of about
the transverse size of the nuclei. The high-$p_T$ parton which was created
in a hard partonic reaction at the earliest stage of nuclear collision is
propagating through this medium. This process ends up with production of a
hadron detected at macroscopic distances. The momenta of the hadrons in
the c.m. frame of collisions ranges from a few up to $10\GeV$ (in the
available data), which overlaps with the kinematics of HERMES
\cite{hermes} and experiment running at Jefferson Lab \cite{will}.

The kinematics and interpretation of this process are much less certain
than in DIS.  In the latter case the initial quark energy is known, as
well as $z_h$, and the density and size of the medium. On the contrary,
the production cross section of a high-$k_T$ parton is a result of
convolution of the initial parton distributions, hard partonic cross
section, and fragmentation functions. Therefore the initial parton energy
and values $z_h$ are not known. Besides, one has to sum over different
species of partons.

These measurements are aimed at probing the density of the created medium,
which is varying in space and time. This brings even more model dependence
to the interpretation of data. In view of these uncertainties in
interpretation of the data on heavy ion collisions, one desperately needs
a reliable model of in-medium hadronization tested in more certain
situations like DIS.

There is a principal difference between hadronization processes in DIS and
high-$p_T$ hadron production at mid rapidities. The dependence of the
production length, $l_p$, on the jet energy varies substantially
depending on the process. Indeed, our previous experience says that the
production time is proportional to the jet energy (the Lorentz factor). In
soft interactions,
 \beq
l_p \,\propto\, \frac{\nu}{\kappa}\ .
\label{2010}
 \eeq
 The string tension $\kappa$ in the denominator provides the correct
dimension for $l_p$.

In the case of DIS the production time is also proportional to energy,
but depends on the photon virtuality,
 \beq
l_p \,\propto\, \frac{\nu}{Q^2}\ .
\label{2020}
 \eeq
 Here the right dimension is provided by the photon virtuality, $Q^2$.
The physics is transparent: the higher is $Q^2$, the stronger is the kick
from the photon, the more intensive is gluon radiation and vacuum energy
loss by the quark. Correspondingly, the shorter must be the color
neutralization time, $l_p$, if one wants to produce a leading hadron with
large $z_h$.

Now we are coming to the point. In the case of $90^o$ parton scattering,
like in the process illustrated in Fig.~\ref{aa}, the jet energy
[$\nu$(DIS)$\Rightarrow k_T$(RHI)] and the parton virtuality
[$Q^2$(DIS)$\Rightarrow k_T^2$(RHI)] are controlled by the same
parameter, which is the transverse momentum of the parton, $k_T$.
Therefore, in this case,
 \beq
l_p \,\propto\,\frac{1}{k_T}\ .
\label{2030}
 \eeq
 It is clear why the energy dependence of the production time switches to
the inverse. The vacuum energy loss with a rate proportional to $k_T^2$
is so intensive, that in spite of the Lorentz factor $k_T$, the
hadronization process must finish shortly after the hard partonic
collision, otherwise the parton energy will degrade too much, making
impossible production of an energetic hadron.

 It is also clear that in pure perturbative QCD calculations the
dimensional parameter, $\Lambda_{QCD}$, can emerge only under a log.
Therefore, Eq.~(\ref{2030}) presents the unique possibility to provide
the correct dimension for $l_p$. Thus, the higher is the energy of the
jet, the faster the leading pre-hadron is produced. This outrageous
conclusion may contradict simple intuition based on the experience with
the string model.

One may think that no pre-hadron can be produced inside a deconfined
medium. This is not correct. Since the mean radius of the produced
pre-hadron decreases like $1/k_T$, it should be smaller that the Debye
screening radius at large $k_T$. Later on, the pre-hadron is evolving
its size and may be dissolved in the medium, but that would just mean a
large absorption cross section.

\begin{figure}[tbh]
 \centerline{\psfig{figure=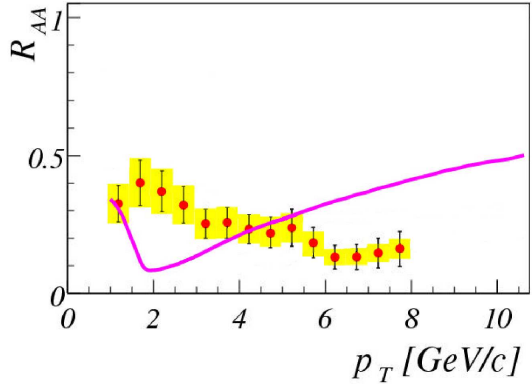,width=7cm}}  
\begin{center}\parbox{13cm}{
 \protect\caption{Data from the PHENIX experiment \cite{qm02} for pion
suppression in $10\%$ central gold-gold collisions at $\sqrt{s}=130\GeV$
in comparison with the prediction from \cite{ww}. The figure is borrowed
from Ref.~\cite{qm02}.}
 \label{ww-data} 
}\end{center}
 \end{figure}

The energy dependence of the production length suggested by
Eq.~(\ref{2030}) for high-$p_T$ processes is inverse to what one has
for DIS, Eq.~(\ref{2020}). Correspondingly, the energy dependence of
nuclear suppression should be opposite to DIS. As long as the nuclear
ratio in DIS rises with energy (Figs.~\ref{pi-nu}--\ref{k-pi-nu}), we
should expect a falling $p_T$-dependence of the nuclear ratio for heavy
ion collisions. This unusual effect was indeed observed at RHIC. Data
from the PHENIX experiment depicted in Fig.~\ref{ww-data} demonstrate
quite a strong fall of the ratio at $p_T>2\GeV$.
 Of course at high $p_T$ one should expect saturation ($l_p=0$) and the
following rise due to color transparency.

On the contrary, in the energy loss scenario, there is no principal
difference between the two cases. As long as the nuclear ratio rises with
energy in DIS, it must rise with $p_T$ in heavy ion collisions (unless one
makes different models of hadronization in the two reactions). Such a
behavior was indeed predicted in \cite{ww} as is shown in
Fig.~\ref{ww-data}. That was a true prediction before the data at high
$p_T$ were released. Of course, once the data are known, one can find an
explanation by introducing new, exotic physical assumptions.

\section{Conclusions and discussion}

The model of in-medium hadronization \cite{knp} developed long ago (last
century) has successfully predicted the nuclear effects for inclusive
electroproduction of leading hadrons. It was realized that the string
model is rather irrelevant to the early stage of hadronization in such a
perturbative QCD process as DIS. Instead, there has been developed a
perturbative description which possessed a clear pattern for the
space-time evolution, and allowed to carry out numerical estimates. The
key points of the model are as follows.

\begin{itemize}
 \item
 A leading quark originated from DIS loses energy for hadronization which
we treat perturbatively via gluon bremsstrahlung. It is important to
discriminate between {\it vacuum} and {\it induced} energy losses
(which are frequently mixed up). The
former is always present, even on a free nucleon target, and is the main
source of energy loss. In the case of in-medium hadronization the quark
radiates more gluons due to multiple soft interaction with the medium.
This correction is usually much smaller than the vacuum energy loss.
 \item
 It is important to discriminate between {\it production} and {\it
formation} times (which are frequently mixed up). Energy loss (both
vacuum and induced) stops when color neutralization occurs, i.e. the
leading quark picks up an antiquark and produces a pre-hadron, which is a
colorless $\bar qq$ dipole with no stationary wave function. The
corresponding time interval is called {\it the production time}, or
production length. Energy conservation enforces the production time to
vanish at the kinematic limit of production of hadrons with maximal
possible energy ($z_h\to 1$).
 \item
 The pre-hadron attenuates on its way out of the nucleus with an
absorptive cross section which is controlled by the varying dipole size.  
Color transparency is an important ingredient of this dynamics.
Eventually, on a longer time scale, called {\it formation time} (or
length) the dipole develops the hadronic wave function. We employ a
rigorous quantum-mechanical description of this process within the
light-cone Green function approach. Contrary to the production length
vanishing at $z_h\to 1$, the formation length reaches a maximal value in
this limit.
 \item
 All these effects important for nuclear modification of the hadron
production rate can be evaluated. Then the model is able to predict in a
parameter free way the nuclear effects in DIS as function of energy,
$z_h$, $Q^2$ and $p_T$. Data from HERMES nicely confirm the predictions.
 \item
 Searching for observables which are able to disentangle different
models, we concluded that the energy loss scenario should have problems
explaining the already available data for flavor and $Q^2$ dependence of
nuclear suppression. This model is insensitive to the hadronic size, but
the difference in fragmentation function at $z_h\to 1$ leads to more
suppression for kaons than for pions. This is at variance with HERMES
data. Also the comparison of data on different reactions at different hard
scales led to a conclusion that $p_T$ broadening and energy loss rise with
$Q^2$. Such an expectation also contradicts HERMES data. It would be very
informative to have data on the variation of the $p_T$-distribution with 
$z_h$,
which is a direct way to measure the production length.
 \item
 The new feature of hadronization dynamics in high-$p_T$ processes is an
inverse dependence of the production time on the jet energy compared to
what is known for DIS. Correspondingly, the $p_T$-dependence of nuclear
suppression in heavy ion collisions should be enhanced at larger $p_T$,
contrary to DIS where nuclear effects vanish at high energy. Such a
puzzling behavior was indeed disclosed by recent measurements at RHIC.
Nevertheless, at higher $p_T$ we expect the effect of color transparency
to take over, then the $k_T$-factorization will be restored.

\end{itemize}

 {\bf Acknowledgments:} We are grateful for inspiring and helpful
discussions to Claudio Ciofi degli Atti, Miklos Gyulassy, Mikkel Johnson,
Larry McLerran, Hans-J\"urgen Pirner, Andreas Sch\"afer, and Ivan Schmidt.
We especially acknowledge the careful reading of the manuscript and
numerous valuable comments made by Alberto Accardi, Will Brooks, Valeria
Muccifora, and Pasquale Di Nezza.  This work was supported by the grant
No.~GSI-OR-SCH from the Gesellschaft f\"ur Schwerionenforschung Darmstadt
(GSI), by the grant INTAS-97-OPEN-31696, by the Slovak Grant Agency VEGA,
and by Alexander von Humboldt Research Fellowship.

\end{document}